\documentclass[longauth]{aa}  
\pdfoutput=1
\usepackage{graphicx}
\usepackage{comment}
\usepackage{hyperref}
\usepackage{txfonts}
\usepackage{xcolor}
\usepackage{epstopdf}   
\def\ms{\hbox{\,m\,s$^{-1}$}}         
\def\m2s2{\hbox{\,m$^{2}$\,s$^{-2}$}} 

\begin{document}

   \title{A sub-Neptune and a non-transiting Neptune-mass companion unveiled by ESPRESSO around the bright late-F dwarf HD 5278 (TOI-130)\thanks{Based on Guaranteed Time Observations collected at the European Southern Observatory under ESO programme 1102.C-0744 by the ESPRESSO Consortium.}}
    \titlerunning{HD 5278 b and c unveiled by ESPRESSO}

\author{A. Sozzetti\inst{1}
  \and M.\,Damasso \inst{1}
  \and A.\,S.\,Bonomo \inst{1}
  \and Y.\,Alibert \inst{2}
  \and S.\,G.\,Sousa \inst{3}
  \and V.\,Adibekyan \inst{3,4}
  \and M.\,R.\,Zapatero Osorio \inst{5}
  \and J.\,I.\,Gonz\'alez Hern\'andez \inst{6,7}
  \and S.\,C.\,C.\,Barros \inst{3,4}
  \and J.\,Lillo-Box \inst{8}
  \and K.\,G.\,Stassun \inst{9}
  \and J.\,Winn \inst{10}
  \and S.\,Cristiani\inst{11,12}
  \and F. Pepe\inst{13}
  \and R.\,Rebolo\inst{6,7,14}
  \and N.\,C.\,Santos\inst{3,4}
  \and R.\,Allart \inst{15,13}
  \and T. Barclay \inst{16,17}
  \and F.\,Bouchy \inst{13}
  \and A.\,Cabral \inst{18,19}
  \and D.\,Ciardi \inst{20}
  \and P.\,Di\,Marcantonio \inst{11}
  \and V.\,D'Odorico \inst{11,12}
  \and D.\, Ehrenreich \inst{13}
  \and M.\,Fasnaugh \inst{21}
  \and P.\,Figueira \inst{3,22}
  \and J.\,Haldemann \inst{2}
  \and J.\,M.\,Jenkins \inst{23}
  \and D.\,W.\,Latham \inst{24}
  \and B.\,Lavie \inst{13}
  \and G.\,Lo Curto \inst{22}
  \and C.\,Lovis \inst{13}
  \and C.\,J .\,A.\,P.\,Martins \inst{3,25}
  \and D.\,M\'egevand \inst{13}
  \and A.\,Mehner \inst{22}
  \and G.\,Micela \inst{26}
  \and P.\,Molaro \inst{6,7}
  \and N.\,J.\,Nunes \inst{18,19}
  \and M.\,Oshagh \inst{6,7}
  \and J. Otegi \inst{27,13}
  \and E.\,Pall\'e \inst{6,7}
  \and E.\,Poretti \inst{28}
  \and G.\,Ricker \inst{29}
  \and D.\,Rodriguez \inst{30}
  \and S.\,Seager \inst{31,32,33}
  \and A.\,Su\'arez Mascare\~no \inst{6,7}
  \and J.\,D.\,Twicken \inst{34}
  \and S.\,Udry \inst{13}
          }
    \authorrunning{A. Sozzetti et al.}
    
   \institute{INAF - Osservatorio Astrofisico di Torino, Via Osservaorio 20, I-10025 Pino Torinese, Italy\\
              \email{alessandro.sozzetti@inaf.it}
\and Physics Institute of University of Bern, Gesellschaftsstrasse\,6, CH-3012 Bern, Switzerland
\and Instituto de Astrof\'isica e Ci\^encias do Espa\c{c}o, Universidade do Porto, CAUP, Rua das Estrelas, 4150-762 Porto, Portugal
\and Departamento de F\'isica e Astronomia, Faculdade de Ci\^encias, Universidade do Porto, Rua do Campo Alegre, 4169-007 Porto, Portugal
\and Centro de Astrobiolog\'{\i}a (CSIC-INTA), Carretera de Ajalvir km 4, E-28850 Torrej\'on de Ardoz, Madrid, Spain
\and Instituto de Astrofisica de Canarias,  Via Lactea, E-38200 La Laguna, Tenerife, Spain
\and Universidad de La Laguna, Departamento de Astrof\'isica, E- 38206 La Laguna, Tenerife, Spain
\and Centro de Astrobiolog\'{\i}a (CSIC-INTA), ESAC campus, E-28692 Villanueva de la Ca\~nada, Madrid, Spain\\ \and Vanderbilt University, Department of Physics \& Astronomy, 6301 Stevenson Center Lane, Nashville, TN 37235, USA
\and Department of Astrophysical Sciences, Princeton University, Princeton, NJ 08544, USA
\and INAF -- Osservatorio Astronomico di Trieste, Via Tiepolo 11, I-34143 Trieste, Italy
\and Institute for Fundamental Physics of the Universe, IFPU, Via Beirut 2, 34151 Grignano, Trieste, Italy
\and D\'epartement d'astronomie, Universit\'e de Gen\`eve, Chemin Pegasi 51, 1290 Versoix, Switzerland
\and Consejo Superior de Investigaciones Cient\'ificas, E-28006 Madrid, Spain
\and Department of Physics, and Institute for Research on Exoplanets, Universit\'e de Montr\'eal, Montr\'eal, H3T 1J4, Canada
\and NASA Goddard Space Flight Center, 8800 Greenbelt Road, Greenbelt, MD 20771, USA
\and University of Maryland, Baltimore County, 1000 Hilltop Circle, Baltimore, MD 21250, USA
\and Instituto de Astrof\'isica e Ci\^encias do Espa\c{c}o, Faculdade de Ci\^encias da Universidade de Lisboa, Campo Grande, PT1749-016 Lisboa, Portugal
\and Departamento de F\'isica da Faculdade de Ci\^encias da Univeridade de Lisboa, Edifício C8, 1749-016 Lisboa, Portugal
\and Caltech/IPAC, 1200 E. California Blvd., Pasadena, CA 91125, USA
\and Department of Physics and Kavli Institute for Astrophysics and Space Research, Massachusetts Institute of Technology, Cambridge, MA 02139, USA
\and ESO, European Southern Observatory, Alonso de Cordova 3107, Vitacura, Santiago
\and NASA Ames Research Center, Moffett Field, CA 94035, USA
\and Center for Astrophysics, Harvard \& Smithsonian, 60 Garden Street, Cambridge, MA 02138, USA
\and Centro de Astrof\'isica da Universidade do Porto, Rua das Estrelas, 4150-762 Porto, Portugal
\and INAF -- Osservatorio Astronomico di Palermo, Piazza del Parlamento 1, 90134 Palermo, Italy
\and Institute for Computational Science, University of Zurich, Winterthurerstr. 190, CH-8057 Zurich, Switzerland
\and INAF -- Osservatorio Astronomico di Brera, Via Bianchi 46, I-23807 Merate, Italy
\and Kavli Institute for Astrophysics and Space Research, Massachusetts Institute of Technology, Cambridge, MA 02139, USA 
\and Space Telescope Science Institute, 3700 San Martin Drive, Baltimore, MD 21218, USA
\and Department of Physics and Kavli Institute for Astrophysics and Space Research, Massachusetts Institute of Technology, Cambridge, MA 02139, USA
\and Department of Earth, Atmospheric and Planetary Sciences, Massachusetts Institute of Technology, Cambridge, MA 02139, USA
\and Department of Aeronautics and Astronautics, MIT, 77 Massachusetts Avenue, Cambridge, MA 02139, USA 
\and SETI Institute, Mountain View, CA 94043, USA
             }

   \date{Received ???; accepted ???}

 
  \abstract
   {Transiting sub-Neptune-type planets, with radii approximately between 2 and 4 R$_\oplus$, are of particular interest, as their study allows us to gain insight on the formation and evolution of a class of planets not found in our Solar System. 
   }
   {We exploit the extreme radial velocity (RV) precision of the ultra-stable echelle spectrograph ESPRESSO on the VLT to unveil the physical properties of the transiting sub-Neptune TOI-130 b, uncovered by the TESS mission orbiting the nearby, bright, late F-type star HD 5278 (TOI-130) with a period $P_{\rm b}=14.3$ days.}
   {We use 43 ESPRESSO high-resolution spectra and broad-band photometry information to derive accurate stellar atmospheric and physical parameters of HD 5278. We exploit the TESS light curve and spectroscopic diagnostics to gauge the impact of stellar activity on the ESPRESSO RVs. We perform separate as well as joint analyses of the TESS photometry and the ESPRESSO RVs using fully Bayesian frameworks to determine the system parameters. }
   {Based on the ESPRESSO spectra, the updated stellar parameters of HD 5278 are T$_\mathrm{eff}=6203\pm64$ K, $\log g =4.50\pm0.11$ dex, [Fe/H]=$-0.12\pm0.04$ dex, M$_\star=1.126_{-0.035}^{+0.036}$ M$_\odot$ and R$_\star=1.194_{-0.016}^{+0.017}$ R$_\odot$. 
   We determine HD 5278 b's mass and radius to be $M_{\rm b} = 7.8_{-1.4}^{+1.5}$ M$_\oplus$ and $R_{\rm b} = 2.45\pm0.05$ R$_\oplus$. The derived mean density, $\varrho_{\rm b} = 2.9_{-0.5}^{+0.6}$ g cm$^{-3}$, is consistent with the bulk composition of a sub-Neptune with a substantial ($\sim30\%$) water mass fraction and with a gas envelope comprising $\sim17\%$ of the measured radius. 
   Given the host brightness and irradiation levels, HD 5278 b is one of the best targets orbiting G-F primaries for follow-up atmospheric characterization measurements with HST and JWST. 
   We discover a second, non-transiting companion in the system, with a period $P_{\rm c}=40.87_{-0.17}^{+0.18}$ days and a minimum mass $M_{\rm c}\sin i_{\rm c} =18.4_{-1.9}^{+1.8}$ M$_\oplus$. 
   We study emerging trends in parameters space (mass, radius, stellar insolation, mean density) of the growing population of transiting sub-Neptunes, and provide statistical evidence for a low occurrence of close-in, $10-15$ M$_\oplus$ companions around G-F primaries with $T_\mathrm{eff}\gtrsim5500$ K. 
   }
   {}


   \maketitle
%

\section{Introduction}

About 27\% of the current catalog of $>2000$ transiting planet candidates uncovered by the TESS mission \citep{Ricker2015} is composed of objects with radii in the range $2.0-4.0$ R$_\oplus$, found in both single and multiple configurations. These objects are usually referred to as sub-Neptunes or mini-Neptunes\footnote{In the remainder of the paper we will adopt the nomenclature sub-Neptunes to refer to this class of planets}. They populate the region at the high-end of the so-called radius valley, a prominent feature of the bimodal radius distribution of small-size ($1-4$ R$_\oplus$), close-in ($P\lesssim 100$ days) planets that was recently unveiled based on the analysis of Kepler mission data \citep{Fulton2017,Fulton2018}. The class of sub-Neptune-type planets does not exist in our Solar System. Understanding why they seem so abundant around other stars (of varied spectral type) in terms of their formation scenarios, evolutionary paths, and range of structural properties is still an open question in exoplanetary science.

Broadly speaking, there is agreement today on the fact that close-in sub-Neptunes cannot be composed of purely rocky material (mixtures of iron and silicates), but must retain more or less significant fractions of volatile elements \citep{Lopez2012,Rogers2015,Lozovsky2018,Jin2018,Bitsch2019,Venturini2020}. However, when it comes to details for explaining the full extent of the radius distribution of these objects, substantially different predictions from theoretical modeling arise. For instance, they could be rocky worlds with small amounts of volatiles in their interiors and substantial, thick H$_2$/He gaseous envelopes of primordial origin (e.g., \citealt{Owen2017,VanEylen2018,Gupta2019}). These go sometimes by the definition of gas dwarfs \citep{Buchhave2014}. Alternatively, they might contain significant amounts of water in primarily solid (ices) form, with thin, H$_2$/He-dominated atmospheres (e.g., \citealt{Leger2004,Zeng2019,Madhusudhan2020,Venturini2020}). In the literature, these are referred to as water worlds or ocean planets. However, irradiated water-rich rocky planets might also possess endogenic thick H$_2$O-dominated, steam atmospheres, in which up to $100\%$ of the planetary water content appears in vaporized form \citep{Dorn2018,Zeng2019,Turbet2020,Mousis2020}. If interactions between the (initially primordial, H$_2$-dominated) atmosphere and magma oceans at the surface are considered, then close-in sub-Neptunes could have atmospheres with variable degrees of (exogenic) hydrogen and (endogenic) water \citep{Kite2020}. 

In principle, density determination for transiting planets with measured radius and mass allows to directly infer their bulk composition. However, mass-radius relationships for small, low-mass planets from theoretical modeling (e.g., \citealt{Bitsch2019,Turbet2020}, and references therein) carry intrinsic degeneracies, with planets of vastly different composition predicted to have observationally  indistinguishable bulk densities. This difficulty is also seen in the case of sub-Neptunes: discriminating between water worlds / ocean planets and rocky planets with a H$_2$–He atmosphere is not possible based on mass and radius determination alone, particularly for those objects with radii in the $2.0-3.0$ R$_\oplus$ range \citep{Adams2008,MIllerRicci2009,Lozovsky2018}. Follow-up atmospheric characterization measurements are therefore necessary. However, density measurements are still of critical importance, as atmospheric analyses rely on knowledge of mass and radius. As a matter of fact, {\it precise} mass determination often remains the limiting factor in atmospheric characterization studies for objects with radii below that of Neptune. Recent work \citep{Batalha2019} advocates for masses measured to better than 20\% precision (i.e. at the $> 5\sigma$ level) in order for constraints on atmospheric composition to be limited solely by the quality of the transmission/emission spectroscopy datasets. 

Achieving statistically significant mass determinations for sub-Neptune-type planets is not an easy task. By design, the TESS mission transit candidates are found orbiting stars significantly brighter than those observed by the Kepler mission, somewhat alleviating the problem of insufficient RV measurement precision due to high photon noise that prevented systematic RV follow-up for the vast majority of the Kepler candidates. However, ground-based follow-up efforts with precision RVs still need to deal with correlated noise sources of stellar origin, typically with amplitudes of the RV variations comparable or exceeding those of the planetary signals. Furthermore, the planetary systems often contain more than one (not necessarily transiting) planet, increasing the complexity of the Keplerian signals and boosting the demands for investment in observing time. Finally, the larger RV amplitudes of the planetary signals make high statistical confidence detections easier to attain around lower-mass primaries. For instance, the masses of sub-Neptunes orbiting stars of any spectral type are measured with median precision of $\sim30\%$, and only a third of the sample has masses determined to better than $20\%$. However, for G-F dwarf primaries (which we operationally define as having $5400\lesssim$T$_\mathrm{eff}\lesssim 7000$ K and $\log g > 4.0$ in the remainder of the paper) the same class of planets has a median precision in mass determination of $\sim40\%$, and only $12\%$ has masses measured at the $>5\sigma$ level, or better\footnote{Data from the transiting exoplanet catalogue TEPCat: \url{https://www.astro.keele.ac.uk/jkt/tepcat/}}. It is therefore desirable to increase the sample of transiting sub-Neptunes with well-determined densities, particularly in the sparsely populated regime of such companions orbiting F-G-type primaries. 

In this paper we present RV measurements from the new ultra-stable spectrograph ESPRESSO (\citealt{Pepe2014,Pepe2020}) that allow us to obtain a  high-precision mass determination for a transiting sub-Neptune candidate with a period of 14.3 days discovered by TESS orbiting the bright ($V=8.0$ mag), high-proper motion late-F dwarf HD 5278 (TIC-263003176, TOI-130, HIP 3911). Our ESPRESSO observations also conclusively reveal the presence of an additional, non-transiting companion with a period of $\sim41$ days. 

In Sect. 2 we present the TESS photometric time-series and the spectroscopic follow-up observations. Sect. 3 focuses on the presentation and discussion of the properties of the host star HD 5278. In Sect. 4 we present our data analysis and results. Sect. 5 contains a discussion on the properties of the HD 5278 planetary system, particularly seen in the context of the sample of presently-known sub-Neptunes orbiting G-F-type primaries. We summarize our results and provide concluding remarks in Sect .6.

\section{Observations}
\subsection{TESS Photometry}

TESS observed TOI-130 (HD 5278) on its Camera 3 at 2-minute cadence in four non-consecutive Sectors: Sector 1 (UT July 25 - August 22 2018), Sector 12 (UT May 21 - June 18 2019), Sector 13 (UT June 19 - July 17 2019), and Sector 27 (UT July 4 - July 30 2020). The target will be re-observed again in Sector 39 (UT May 26 - June 24 2021). The publicly available TESS photometric data were downloaded from the Mikulski Archive for Space Telescopes (MAST) portal\footnote{\url{mast.stsci.edu/portal/Mashup/Clients/Mast/Portal.html}}. We used the light curve produced by the Presearch Data Conditioning Simple Aperture Photometry (PDCSAP; \citealt{Smith2012,Stumpe2012,Stumpe2014}) algorithm adopted for processing the TESS images by the NASA Ames Science Processing Operations Center (SPOC; \citealt{Jenkins2016}). The initial identification of TOI-130.01 as planet candidate by SPOC was obtained in several steps, including a transit search using the Transiting Planet Search Module (TPS; \citealt{Jenkins2002,Jenkins2010}), and the evaluation of a set of transiting planet model fits and internal validation tests \citep{Twicken2018,Li2019}. The reported period of the planet candidate based on the Sectors 1-13 transit search was 14.339 days, with a preliminary transit depth of $409\pm17$ ppm. A total of 8 transits were identified, two in each TESS sector, and the full dataset will be used in the following analysis.

\subsection{Reconnaissance spectroscopy with CORALIE}

We started pursuing the confirmation of the planet candidate collecting spectroscopic observations with the CORALIE spectrograph \citep{Queloz2000} mounted on the 1.2 m Swiss telescope at La Silla Observatory. A total of 14 spectra were obtained between UT September 7 2018 and UT October 5 2018, spanning approximately two orbit cycles. Exposure time was set to 900 sec. Radial velocities were derived with the automated CORALIE pipeline \citep{Queloz2000,Segransan2010}. The rms of the dataset is 10.5 m s$^{-1}$, comparing favorably to the mean internal error of the observations (7.5 m s$^{-1}$) and conclusively ruling out the possibility of an eclipsing binary, that would otherwise produce a signal with a much higher RV amplitude.

\subsection{Spectroscopic follow-up with ESPRESSO}

We carried out the RV monitoring of HD 5278 with ESPRESSO within the context of the Guaranteed Time Observations (GTO) sub-program aimed at measuring precise masses of transiting planet candidates uncovered by the TESS and K2 missions (Program ID 1102.C-744, PI: F.Pepe). HD 5278 was observed over a total time span of 403 days between October 2018 and December 2019. All observations were carried out using the Fabry Perot (FP) as simultaneous calibration source, enabling the monitoring of instrumental drifts down to the 10 cm s$^{-1}$ precision level for which ESPRESSO has ultimately been designed. The complete time series encompasses 43 high-resolution spectra, gathered with the 1-UT (UT3) mode at $R=138\,000$ and acquired with a fixed exposure time of 900 sec. The spectra have a median S/N$\simeq230$ per pixel at $\lambda= 5500$ \AA\, . 

Radial velocities were extracted using version 2.0.0 of the ESPRESSO data reduction software (DRS), which is available for direct download from the ESO pipelines website \footnote{\url{http://www.eso.org/sci/software/pipelines/}}. The DRS (Lovis et al. in prep.) outputs RV measurements based on a Gaussian fit of the cross correlation function (CCF) of the spectrum using a binary mask computed from a stellar template \citep{Baranne1996,Pepe2000}. RV information is extracted from the full wavelength range covered by the instrument ($3800 - 7880$ \AA). For HD 5278, an F9V mask was utilized. The resulting RV time-series has a median internal error of 38 cm s$^{-1}$. From the DRS-extracted ESPRESSO spectra we also obtained (through the DACE interface\footnote{The Data Analysis Center for Exoplanets (DACE) platform is available at \url{https://dace.unige.ch}}) the time-series of a number of useful spectroscopic diagnostics of stellar activity, i.e. full-width half maximum (FWHM), line bisector span (BIS), Mount Wilson $S$-index ($S_\mathrm{MW}$), $\log R^\prime_\mathrm{HK}$, and H$\alpha$ index. The RVs and activity indicators (along side their formal uncertainties) used in this work are listed in Tables \ref{tab:rv} and \ref{tab:activ}, respectively. 


%

\section{Stellar Properties}\label{starprop}

\begin{figure}
    \centering
        \includegraphics[width=.49\textwidth]{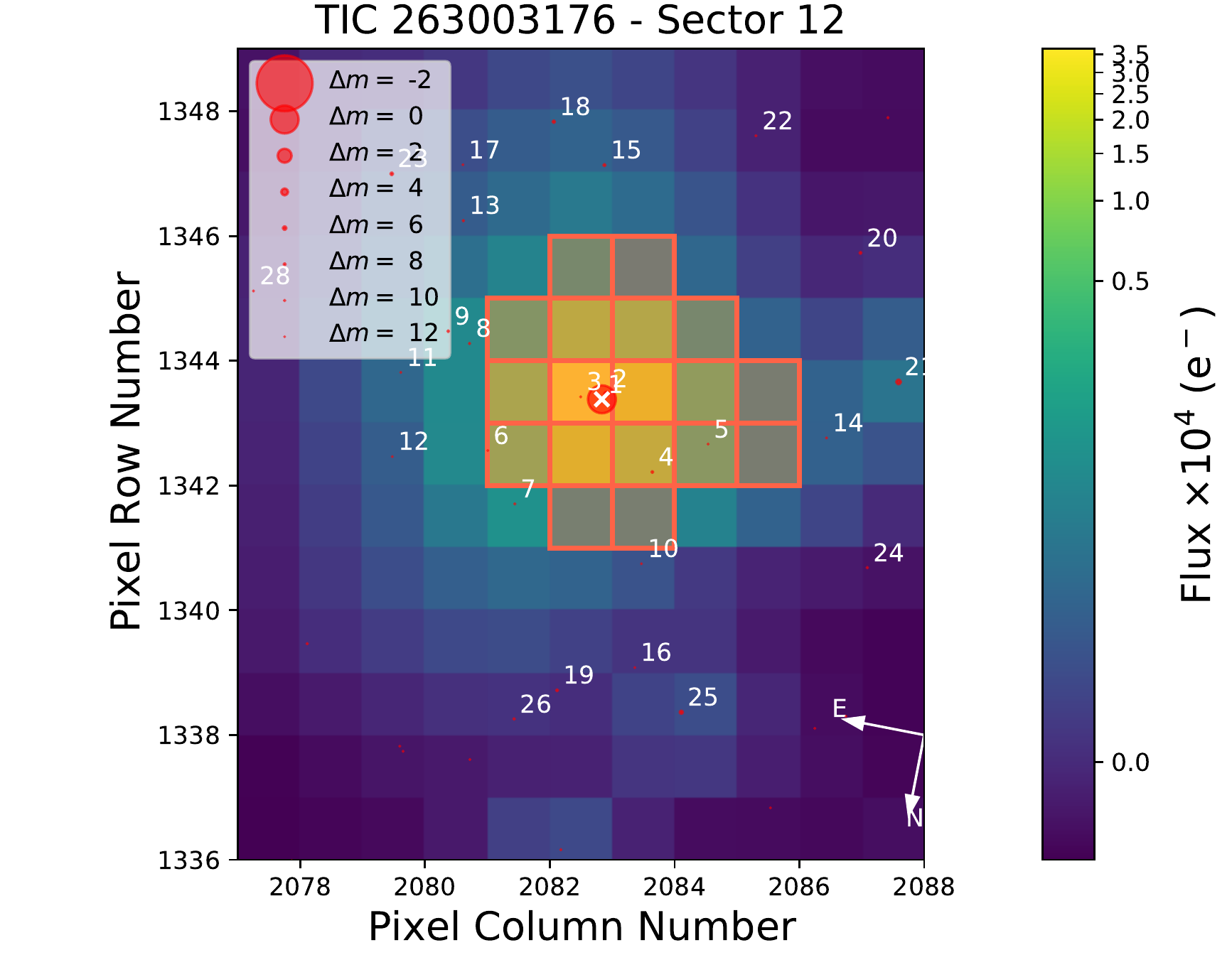}
\caption{Target Pixel File (TPF) of HD 5278 from the TESS observations in Sector 12 (composed with \texttt{tpfplotter}, \citealt{Aller2020}). The SPOC pipeline aperture is overplotted with a red shaded region and the Gaia DR2 catalog is also overlaid with symbol sizes proportional to the magnitude contrast with the target, which is marked with a white cross.}
    \label{fig:TPF}
\end{figure}

\begin{table}
   \caption[]{Astrometry, photometry, and spectroscopically derived stellar properties of HD 5278}
          \label{tab:starprop}
          \centering
         \small
    \begin{tabular}{l c c}
             \hline
             \hline
             \noalign{\smallskip}
             \multicolumn{3}{c}{{\it HD 5278, TOI-130, TIC 263003176, HIP 3911, TYC 9490-786-1}}  \\
             \noalign{\smallskip}
             Parameter     &  Value & Refs. \\
             \noalign{\smallskip}
             \hline
             \noalign{\smallskip}
             \noalign{\smallskip}
             \textit{Astrometry:} \\
             \noalign{\smallskip}
             $\alpha$ (J2000) & 00:50:09.90 & [1,2]  \\
             \noalign{\smallskip}
             $\delta$ (J2000) & $-$83:44:38.03 & [1,2]  \\
             \noalign{\smallskip}
             $\mu_\alpha$ [mas yr$^{-1}$] & $139.42\pm0.05$ & [1,2]  \\
             \noalign{\smallskip}
             $\mu_\delta$ [mas yr$^{-1}$] & $30.47\pm0.04$ & [1,2]  \\
             \noalign{\smallskip}
             $\varpi$ [mas] & $17.33\pm0.03$ & [1,2]   \\
             \noalign{\smallskip}
             $d$ [pc] & $56.12_{-0.81}^{+0.83}$ & [3]  \\
             \noalign{\smallskip}
             \noalign{\smallskip}
             \textit{Photometry:} \\
             \noalign{\smallskip}
             $FUV_\mathrm{GALEX}$ &  $18.955\pm0.080$ & [4] \\
             \noalign{\smallskip}
             $NUV_\mathrm{GALEX}$ &  $12.687\pm0.002$ & [4]\\
             \noalign{\smallskip}
             $B_T$ & $8.563\pm0.016$ & [5]  \\
             \noalign{\smallskip}
             $V_T$ &  $8.007\pm0.011$ & [5]\\
             \noalign{\smallskip}
             $G$ &  $7.8037\pm0.0003$ & [1,6] \\
             \noalign{\smallskip}
             $J$  & $6.949\pm0.027$ & [7] \\
             \noalign{\smallskip}
             $H$   & $6.673\pm0.031$ & [7] \\
             \noalign{\smallskip}
             $K_s$ &  $6.605\pm0.027$ &  [7]\\
             \noalign{\smallskip}
             $W1$ &  $6.553\pm0.069$ & [8] \\
             \noalign{\smallskip}
             $W2$ &  $6.548\pm0.022$ & [8]\\
             \noalign{\smallskip}
             $W3$ &  $6.575\pm0.015$ & [8]\\
             \noalign{\smallskip}
             $W4$ &  $6.511\pm0.048$ & [8]  \\
             \noalign{\smallskip}
             \noalign{\smallskip}
             \textit{Stellar Parameters:} \\
             \noalign{\smallskip}
             $T_\mathrm{eff}$ [K] & $6203\pm64$ &  [9] \\
             \noalign{\smallskip}
             $\log g$ [dex] & $4.50\pm0.11$ &  [9] \\
             \noalign{\smallskip}
             [Fe/H] [dex] & $-0.12\pm0.04$ &  [9] \\
             \noalign{\smallskip}
             [Mg/H] [dex] & $-0.13\pm0.08$ &  [9] \\
             \noalign{\smallskip}
             [Si/H] [dex] & $-0.13\pm0.04$ &  [9] \\
             \noalign{\smallskip}
             $\xi$ [km s$^{-1}$] & $1.31\pm0.03$ & [9]  \\
             \noalign{\smallskip}
             $M_\star$ [M$_\odot$] & $1.126_{+0.036}^{-0.035}$ &  [9] \\
             \noalign{\smallskip}
             $R_\star$ [R$_\odot$] & $1.194_{+0.017}^{-0.016}$  & [9]  \\
             \noalign{\smallskip}
             $\varrho_\star$ [g cm$^{-3}$] & $0.933_{+0.055}^{-0.053}$  & [9]  \\
             \noalign{\smallskip}
             $L_\star$ [L$_\odot$] & $1.907_{+0.065}^{-0.052}$ & [9]  \\
             \noalign{\smallskip}
             $v\sin i$ [km s$^{-1}$] & $4.1\pm0.5$ &  [9] \\
             \noalign{\smallskip}
             $P_\mathrm{rot}$ [days] & $<16.8$ & [9]  \\
             \noalign{\smallskip}
             $<\log R^\prime_{HK}>$  & $-4.86\pm0.03$  &  [9] \\
             \noalign{\smallskip}
             $t$ [Gyr] & $3.0\pm1.1$ & [9]  \\
             \noalign{\smallskip}
             [$U$,$V$,$W$] [km s$^{-1}$] & [$-48.23\pm0.10$,$4.24\pm0.13$,$9.83\pm0.10$] & [9]  \\
             \noalign{\smallskip}
             \hline
      \end{tabular}
\tablebib{[1] \citealt{Brown2018}; [2] \citep{Lindegren2018}; [3] \citep{BailerJones2018}; [4] \citep{Bianchi2017}; [5] \citep{Hog2000}; [6] \citep{Evans2018}; [7] \citep{Cutri2003}; [8] \citep{Cutri2014}; [9] this work.}
\end{table}
HD 5278 (TOI-130, TIC 263003176, HIP 3911) is a high-proper motion, bright ($V=8.0$ mag, $G=7.8$ mag) late F-type star located at $\sim56$ pc from the Sun \citep{Brown2018,Lindegren2018}. The field around HD 5278 is shown in Figure \ref{fig:TPF}, which reproduces a sample image from the TESS target pixel files of Sector 12 obtained with the publicly available \texttt{tpfplotter}\footnote{\url{https://github.com/jlillo/tpfplotter}} tool \citep{Aller2020}. The number of detected Gaia sources is not very high, and they are all rather faint ($\Delta G > 5.5$ mag), resulting in a relatively low degree of blending. However, the closest and brightest source to HD 5278, with $G= 13.2$ mag, falls well inside the central target pixel. Gaia DR2 reports it at a separation of $\sim 2.2$ arcsec, and with similar parallax and proper motion components with respect to HD 5278, albeit with a high value of the renormalized unit weight error (RUWE = 11.2) due to the presence of the nearby bright target. We used standard criteria to establish their physical association, requiring small differences in parallax and proper motion ($\Delta\varpi < \mathrm{max}[1.0, 3\sigma_\varpi]$, $\Delta\mu < 0.1\mu$; see \citealt{Smart2019}). Both conditions are simultaneously satisfied, therefore the two stars are recognized as a resolved double star system. The Starhorse catalog \citep{Anders2019} reports effective temperature $T_\mathrm{eff} = 3341\pm300$ K and mass $M=0.30_{-0.11}^{+0.14}$ M$_\odot$ for the secondary (Gaia DR2 Source Id 4617759518796452608). 

We summarize in Table \ref{tab:starprop} the main astrometric, photometric, and physical stellar parameters for HD 5278. The stellar atmospheric parameters for the target (effective temperature $T_\mathrm{eff}$, surface gravity $\log g$, microturbulence $\xi$, and iron abundance [Fe/H]) were obtained using a standard technique of excitation and ionization balance (see, e.g., \citealt{Sozzetti2004,Sozzetti2007,Sousa2011}, and references therein). 
To this end, we used the StarII workflow of the data analysis software (DAS) of ESPRESSO \citealt{DiMarcantonio2018} to produce a high-quality 1D ESPRESSO spectrum of HD 5278. We initially coadded, and normalized order by order 43 blaze-corrected bi-dimensional ESPRESSO spectra at the barycentric reference frame. Then these spectra were merged and corrected for RV. The final 1D ESPRESSO spectrum with a pixel size of 0.5 km s$^{-1}$ (see Fig. \ref{1d_spectrum}) has S/N of about 1000 and 2000 at 4200 and 5000 \AA, respectively, and higher than 2000 for longer wavelengths.
Next, the \texttt{ARES v2} code \citep{Sousa2007,Sousa2015} with the \citet{Sousa2008} input linelist was used to consistently measure the equivalent widths (EWs) for each line. Metal abundances were derived under the assumption of local thermodynamic equilibrium (LTE), using
the 2014 version of the spectral synthesis code \texttt{MOOG}\footnote{\url{https://www.as.utexas.edu/~chris/moog.html}} \citep{Sneden1973} and a grid of Kurucz ATLAS plane-parallel model stellar atmospheres \citep{Kurucz1993}. We obtained $T_\mathrm{eff} = 6203\pm64$ K, $\log g = 4.50\pm0.11$, $\xi = 1.31\pm0.03$ km s$^{-1}$, and [Fe/H] $= -0.12\pm0.04$. The quoted uncertainties in the atmospheric parameters include contributions (added in quadrature) from possible systematic errors which amount to 60 K, 0.04 dex, and 0.1 dex for $T_\mathrm{eff}$, [Fe/H], and $\log g$, respectively \citep{Sousa2011}. Using on a classical curve-of-growth analysis method based on the same tools and model atmospheres used for the stellar parameter determination \citep[e.g.][]{Adibekyan2012,Adibekyan2015}, we also report in Table \ref{tab:starprop} our measurements of the Mg and Si abundances in HD 5278, which will be input to the analysis carried out in Sect. \ref{hd5278b_comp}.

\begin{figure}
    \centering
    \includegraphics[width=.35\textwidth, angle=90]{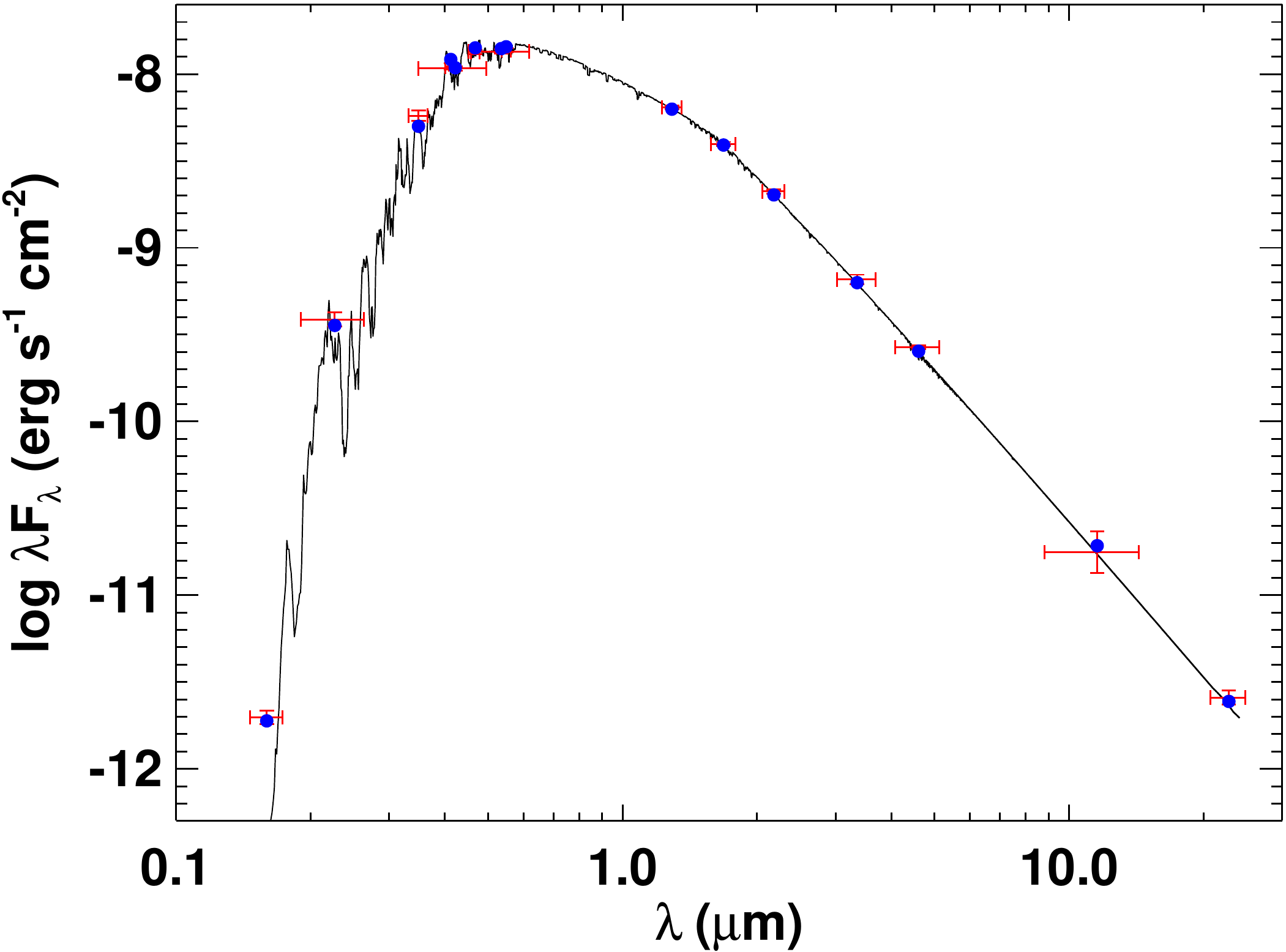}
    \caption{The spectral energy distribution of HD 5278. Red markers depict the photometric measurements with vertical error bars corresponding to the reported measurement uncertainties from the catalog photometry. Horizontal error bars depict the effective width of each passband. The black curve corresponds to the most likely stellar atmosphere model. Blue circles depict the model fluxes over each passband.}
    \label{fig:sedfit}
\end{figure}

\begin{figure*}
    \centering
    \includegraphics[width=0.98\textwidth, angle=0]{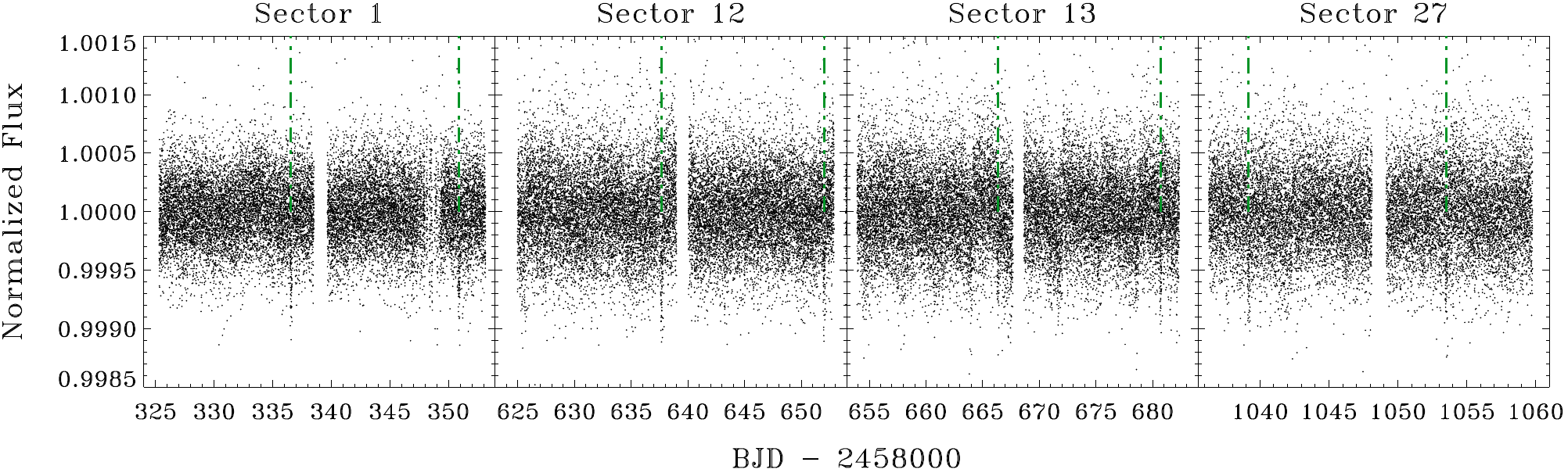}
    \caption{The TESS light curve of HD 5278 from sectors 1, 12, 13, and 27. The eight SPOC-detected transits are highlighted with vertical green dashed-dotted lines throughout.}
    \label{fig:LC}
\end{figure*}

We derived stellar mass, radius, and age using a twofold approach. We first fed the $T_\mathrm{eff}$ and [Fe/H] estimates to the optimization code \texttt{PARAM v1.3} \citep{daSilva2006,Rodrigues2014,Rodrigues2017}, with the additional information of the Gaia DR2 parallax and $V_T$-band magnitude. We obtained $M_\star= 1.103\pm 0.033$ M$_\odot$, $R_\star= 1.148\pm 0.027$ R$_\odot$, and $t=2.2\pm1.4$ Gyr. Then, we used the \texttt{EXOFASTv2} code (see \citealt{Eastman2017,Eastman2019}, for a full description) to perform a fit to the star's broadband spectral energy distribution (SED) from the near ultraviolet to the mid-infrared (see Figure \ref{fig:sedfit}). The SED fit (see e.g. \citealt{Stassun2016}) was carried out using retrieved broadband NUV photometry from GALEX, Tycho-2 B- and V-band magnitudes, Str\"omgren {\it uvby} photometry, 2MASS JHKs near-IR magnitudes, and WISE W1-W4 IR magnitudes and invoking within \texttt{EXOFASTv2} the YY-isochrones \citep{Yi2001} to model the star. We obtained: $M_\star= 1.126_{+0.036}^{-0.035}$ M$_\odot$, $R_\star= 1.194_{+0.017}^{-0.016}$ R$_\odot$, and $t=3.0\pm1.0$ Gyr. With consistent numbers obtained by the two approaches, we elect to list in Table \ref{tab:starprop} the values from the \texttt{EXOFASTv2} analysis that makes use of the well-sampled stellar SED. 

 We computed the Galactic space velocity vector [$U$, $V$, $W$] for HD\,5278 using the radial velocity from Gaia DR2 (RV = $-30.30 \pm 0.18$ km\,s$^{-1}$, \citealt{Brown2018}) and the parallax and proper motion information listed in Table 1. We calculated the heliocentric velocity components in the directions of the Galactic center, Galactic rotation, and north Galactic pole, respectively, using the formulation developed by \citet{Johnson1987}. The right-handed system was employed and the Solar motion was not subtracted from our calculations. The uncertainties associated with each space velocity component were obtained from the observational quantities after the prescription of \citealt{Johnson1987}. 

According to the kinematic criteria defined by e.g., \citealt{Gagne2018} and \citealt{Reddy2006}, HD\,5278 is not a member of any known young stellar moving group with ages below a few hundred Myr,  and it has a 99.9\%~probability of being a member of the field. Its low $V$ and $W$ velocities indicate that HD\,5278 is kinematically a member of the thin disk of the Galaxy for which ages typically younger than 7–9 Gyr are expected \citep{Kilic2017}. This agrees with the intermediate age of a few Gyr derived for this star using the \texttt{EXOFASTv2} tool.

The average value of $\log R^\prime_{HK}$ ($-4.86\pm0.03$) and the $B-V$ colour for HD 5278 imply a predicted stellar rotation period $P_\mathrm{rot} = 9.6\pm1.8$ days \citep{Noyes1984}. This number is compatible with the value obtained from the \citet{Mamajek2008} relations, which predict $P_\mathrm{rot} = 12.6\pm 0.6$ days, and which also predict a gyrochronological age of $4.1\pm0.5$ Gyr, consistent with the estimate in Table \ref{tab:starprop}. We further constrained the value of $P_\mathrm{rot}$ through a direct measurement of the projected rotational velocity ($v\sin i$) of HD 5278 from the ESPRESSO co-added spectrum. The $v\sin i$ was derived based on the analysis of 16 isolated iron lines with the FASMA spectral synthesis package \citep{Tsantaki2018}, fixing all the stellar atmospheric parameters (from Table \ref{tab:starprop}), macroturbulent velocity, and limb-darkening coefficient. The limb-darkening coefficient (0.62) was determined using the aforementioned stellar parameters as described in \citet{Espinoza2015} assuming a linear limb darkening law. The macroturbulent velocity (4.4 km/s) was determined by using the temperature- and gravity-dependent empirical formula from \citet{Doyle2014}. We obtained $v\sin i = 4.1\pm 0.5$ km s$^{-1}$. This allows us to put a $1-\sigma$ upper limit to the stellar rotation period $P_\mathrm{rot} < 16.8$ days (given the stellar radius in Table \ref{tab:starprop}). 

It must be noted how the $v\sin i$ value obtained for HD 5278 is on the low side for a star of this spectral type. For instance, \citet{Winn2017} presented a collection of $v\sin i$ measurements of Kepler stars with transiting planets. There is an overall tendency for $v\sin i$ to rise with effective temperature, as expected (see their Fig. 6). Within the range from 6150 to 6250 K, out of a sample of 42 stars only 2 have $v\sin i$ values as low as 4.1 km s$^{-1}$ (the value obtained for HD 5278).  Thus, either HD 5278 is among the slowest 5-10\% of rotators in its temperature/mass range, or it has a value of $\sin i$ significantly lower than unity.

\section{Data Analysis and System Parameters}

In order to maximize the robustness of the recovered system parameters, we conduct a two-step analysis of our data. In Section \ref{TESSLCanalysis} we first model separately the TESS light curve, with the resulting planet parameters being used as priors in the subsequent analysis of the ESPRESSO RVs in Section \ref{ESPRESSORVanalysis}. In Section \ref{combinedESPRESSOTESS} we describe the global analysis of the combined TESS and ESPRESSO datasets, which we ultimately adopt as fiducial. 

\subsection{TESS transit analysis}\label{TESSLCanalysis}

The \texttt{PDCSAP} light-curve for the four TESS Sectors is shown in Figure \ref{fig:LC}. No clear activity features with amplitude larger than
$\sim300$ ppm are seen in the TESS data, indicating that the host
star shows rather low levels of magnetic activity. Indeed, in the Kepler sample stars of similar $T_\mathrm{eff}$ to that of HD 5278 have amplitudes of the detected periodic variability on the order of 1 mmag, 300 ppm corresponding to the lowest-amplitude periodicities found in the Kepler light-curves (see Figure 3 in \citealt{mcQuillan2014}). We performed an independent transit search using a median filter and the BLS algorithm \citep{Kovacs2002,Bonomo2012}. We could recover the eight transits of TOI-130 b with a period of 14.34 days, but no other potential transit signal was identified. 

For the purpose of the analysis of the TESS \texttt{PDCSAP} light-curve, we  extracted data sub-samples centred around the mid-transit times and extending 1.5 times the transit duration before their ingress and after their
egress, and normalized each transit with a first-order polynomial to the out-of-transit data. The transit parameters were measured based on a Bayesian analysis of the TESS data performed using a differential evolution Markov Chain Monte Carlo (DE-MCMC) method \citep{TerBraak2006,Eastman2013}. 
The free parameters of our model are (under the assumption of a circular orbit) the transit epoch $T_{\rm 0}$, the orbital period $P$, the transit duration $T_{\rm 14}$, the ratio of the planet to stellar radii $R_{\rm p}/R_{\star}$, and the inclination $i$ between the orbital plane and the plane of the sky. As can be seen in Figure \ref{fig:LC}, the quality of the photometric data appears to degrade somewhat in Sectors 12, 13, and 27, with larger uncorrected systematics of likely instrumental nature. We then included in our global model four  scalar jitter terms $\sigma_\mathrm{S_i}$, one for each of the four TESS sectors. The two limb-darkening coefficients (LDC) $u_{1}$ and $u_{2}$ of the quadratic limb-darkening law (e.g., \citealt{Claret2018}) appropriate for the atmospheric parameters of HD 5278 reported in Table \ref{tab:starprop} were initially kept fixed. 
Uninformative priors were used for all model parameters. A Gaussian prior was imposed on the transit stellar density $\varrho_\star$ from the EXOFASTv2 stellar parameters, i.e. $\mathcal{N}$(0.933,0.055) g cm$^{-3}$. This prior indirectly affects $R_p/R_\star$, $i$, and $T_{\rm 14}$, given that these parameters are related to $a/R_\star$, the semi-major axis to stellar radius ratio, through Eq. 12 in \citet{Gimenez2006}, and that $\varrho_\star \propto (a/R_\star)^3$.

A DE-MCMC run with a number of chains equal to twice the number of free parameters was then carried out. After removing the ``burn-in'' steps and 
achieving convergence and good mixing of the chains following the same criteria as in \citet{Eastman2013}, the medians of the posterior distributions and their $\pm34.13\%$ intervals were evaluated and were taken as the final parameters and 
associated $1~\sigma$ uncertainties, respectively. A DE-MCMC run with the inclusion of the LDCs as free model parameters produced indistinguishable results (well within the $1\sigma$ uncertainties on the model parameters). This is not unexpected, partly because of the small number of transit events available, partly because we are in a configuration ($T_\mathrm{eff}> 6000$ K, non-grazing transit) for which the radius ratio is not particularly affected by the choice of fixed LDCs (e.g., \citealt{Csizmadia2013}). The results reported in Table \ref{tab:planet_param} are those obtained with fixed LDCs. 

\subsection{ESPRESSO RV and activity analysis}\label{ESPRESSORVanalysis}

\begin{figure}
    \centering
    \includegraphics[width=0.49\textwidth, angle=0]{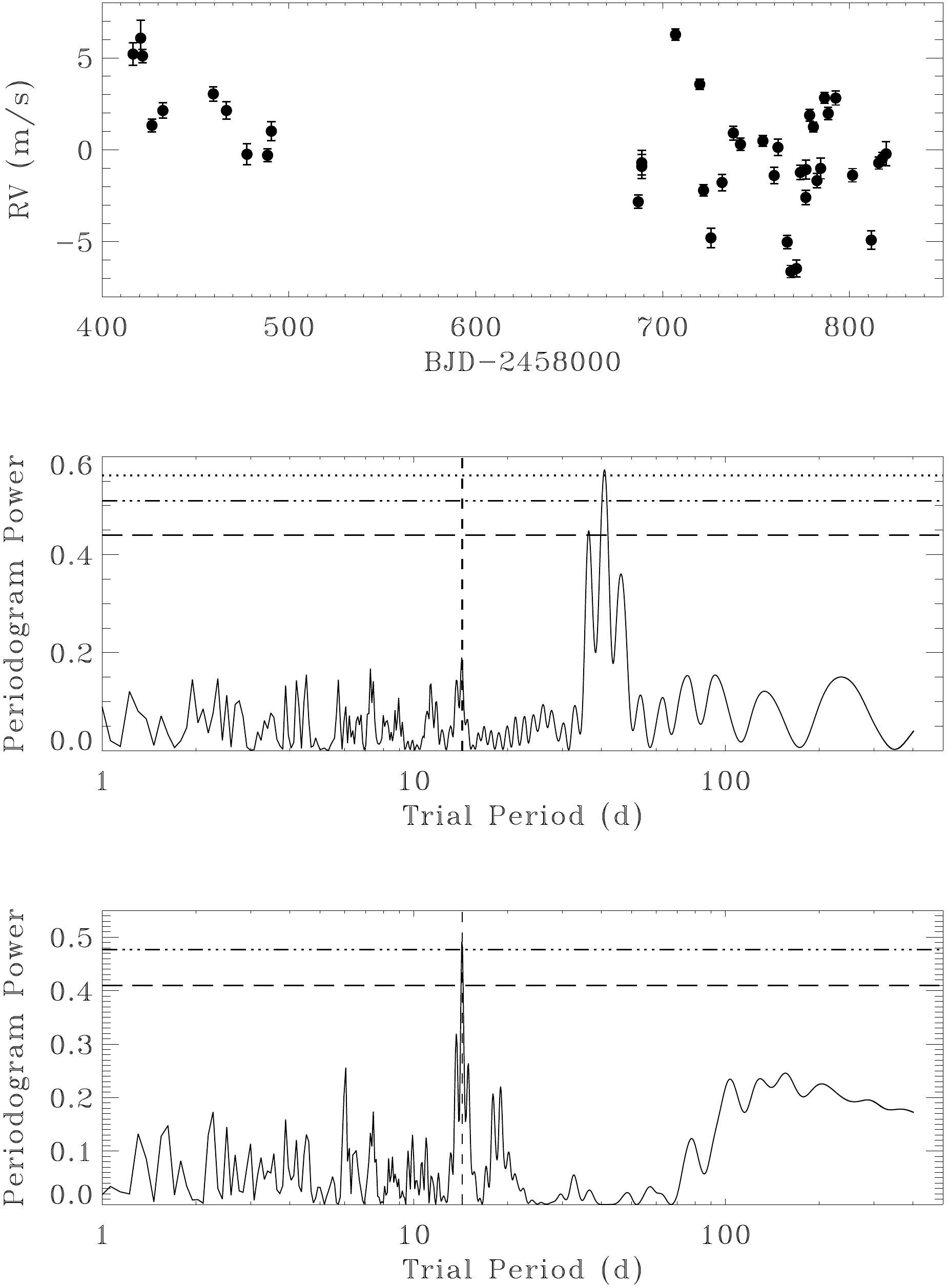}
\caption{Top: time series of the ESPRESSO RVs for HD 5278. Center: GLS periodogram of the original RV time series. Bottom: GLS periodogram of the residuals after removal of a sinusoid with a 41-day period. The vertical dashed line in the central and lower panels shows the period of the transiting planet candidate from TESS photometry. In the center and bottom panels the horizontal long-dashed, dashed-dotted, and dotted lines represent 10\%, 1\%, and 0.1\% FAP levels, respectively.}
    \label{fig:rvgls}
\end{figure}

\begin{figure}
    \centering
    \includegraphics[width=0.49\textwidth, angle=0]{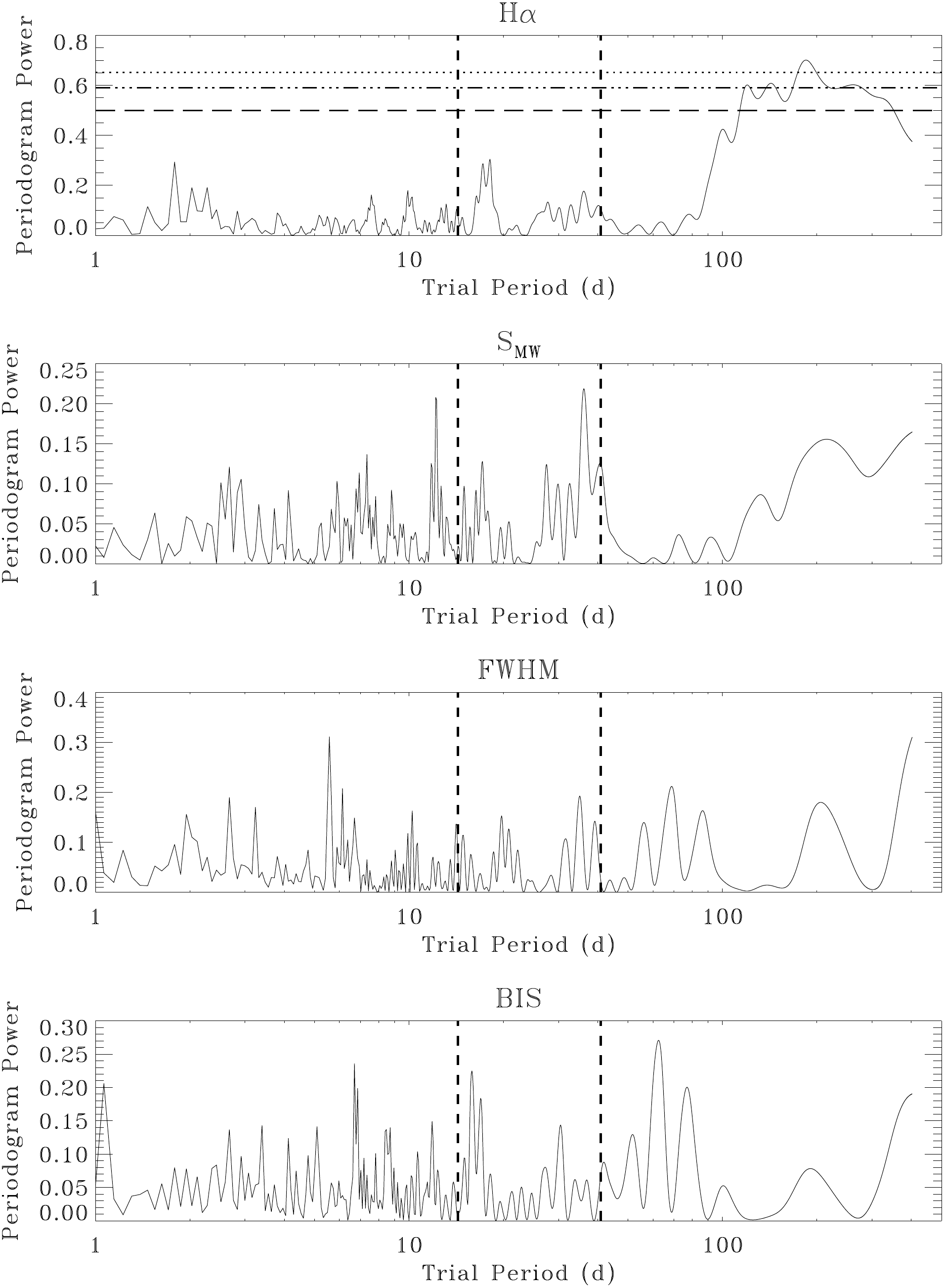}
    \caption{GLS periodograms of various activity diagnostics measured in the ESPRESSO spectra of HD 5278. From the top to bottom: H$\alpha$-index, Mount Wilson S-index, FWHM, and bisector span. In each panel, the two vertical dashed lines show the position of the two periodic signals identified in the RV time series. In the top panel the horizontal lines have the same meaning as in Figure \ref{fig:rvgls}.
    }
    \label{fig:activity}
\end{figure}

The top panel of Figure \ref{fig:rvgls} shows the ESPRESSO RV time-series of HD 5278. The rms of the dataset is 3.1 m/s, almost a factor of 10 larger than the typical internal errors of the observations. A Generalized Lomb-Scargle (GLS; \citealt{Zechmeister2009}) analysis returns a dominant peak at $\sim41$ days (central panel of Figure \ref{fig:rvgls}), with a false alarm probability (FAP) of $0.05\%$, calculated via a standard bootstrap with replacement procedure. At the period of the transiting candidate a secondary peak is visible, and this becomes the dominant periodicity (FAP = $0.66\%$) when GLS is run on the RV residuals after subtraction of a sinusoid with $P=41$ days and amplitude $3.2$ m/s (bottom panel of Fig. \ref{fig:rvgls}). Some structure with modest power at longer periods remains. 

The dominant period in the RV data does not seem to be connected to the rotation period of the star, constrained to be on the order of the orbital period of the transiting companion, or shorter. Further evidence for the Keplerian origin of the signal at $\sim41$ days comes from investigation of the stellar activity indicators. The four panels of Figure \ref{fig:activity} show the GLS periodograms for FWHM, BIS, S-index, and H$\alpha$ index. No peak with significant power is seen in the vicinity of either $P=14.34$ days or $P=41$ days. In fact, no strong evidence of rotation is found, with all periodogram peaks in BIS, FWHM, and S-index reported with FAP$>10\%$. The only significant feature (FAP $< 0.1\%$) is found in correspondence of a long-term trend in the H$\alpha$ index, peaking at around $180$ days. A marginally significant Spearman's rank correlation coefficient (0.30) is only obtained between RVs and H$\alpha$ index. The results from this analysis reinforce the hypothesis that the signal at $41$ days is produced by another, possibly non-transiting, planet in the system, and we therefore proceed to fit a two-Keplerian model to the ESPRESSO RV data\footnote{We did check that the CORALIE RV measurements are not inconsistent with the orbit fitting results presented in this Section but, as expected, they do not have sufficient precision to add useful constraints on the model parameters. }. 

The free parameters of our two-planet model are the time of transit center $T_{c,b}$, the time of inferior conjunction $T_{c,c}$ the orbital periods $P_b$ and $P_c$, two RV zero points $\gamma_1$ and $\gamma_2$ for ESPRESSO data obtained before and after the fiber-link upgrade intervention in Summer 2019, the RV semi-amplitudes $K_b$, $K_c$, $\sqrt{e_b}\cos\omega_b$, $\sqrt{e_b}\sin\omega_b$, $\sqrt{e_c}\cos\omega_c$, and $\sqrt{e_c}\sin\omega_c$, where $e_b$, $e_c$ are the eccentricities and $\omega_b$, $\omega_c$ the arguments of periastron. The lack of evidence in the photometry, RVs, and spectroscopic activity indicators of any relevant modulation around the expected rotation period of the star makes us opt for modeling any additional source of noise in the RVs only in terms of two uncorrelated jitter terms ($\sigma_\mathrm{jit,1}$, $\sigma_\mathrm{jit,2}$) added in quadrature to the formal uncertainties of the RVs before and after the intervention. 

The ESPRESSO RV time-series analysis was carried out using the publicly
available Monte Carlo sampler and Bayesian inference tool
\texttt{MultiNestv3.10} (e.g. \citealt{Feroz2013}), through the 
\texttt{pyMultiNest} wrapper \citep{Buchner2014}. This multimodal nested sampling algorithm was setup to run with 500 live points and a sampling efficiency of 0.3. The results of the analysis are reported in Table \ref{tab:planet_param}. 

The full solution indicates that the possible offset between the two pre- and post-upgrade ESPRESSO datasets is at the level of 1 m s$^{-1}$, similar to that shown by \citet{SuarezMascareno2020} for a star of later spectral type (Proxima Cen). The uncorrelated stellar jitter appears marginally larger during the second season of ESPRESSO observations, but the difference is only at the $1.3\sigma$ level. While the lack of variability in the TESS photometry and spectroscopic activity indicators prevented us from obtaining clear information on the possible rotation period of the primary, we did attempt to employ a more sophisticated modeling framework based on Gaussian Process (GP) regression to mitigate stellar activity effects. We performed a fit to the ESPRESSO RVs using a two-Keplerian + GP with quasi-periodic kernel (for details see e.g., \citealt{Damasso2018,Damasso2020}), with the upper limit on the putative rotation period of the star constrained by the value reported in Table \ref{tab:starprop}. The results (not shown) were unsatisfactory (an outcome not unexpected also considering the relatively small size of the RV dataset), with the rotation parameter unconstrained and a statistically disfavoured model (marginal likelihood $\ln\mathcal{Z}=-97.1$ and $\ln\mathcal{Z}=-98.6$ for the models without and with GP, respectively). 

\begin{figure}
    \centering
    \includegraphics[width=0.45\textwidth, angle=0]{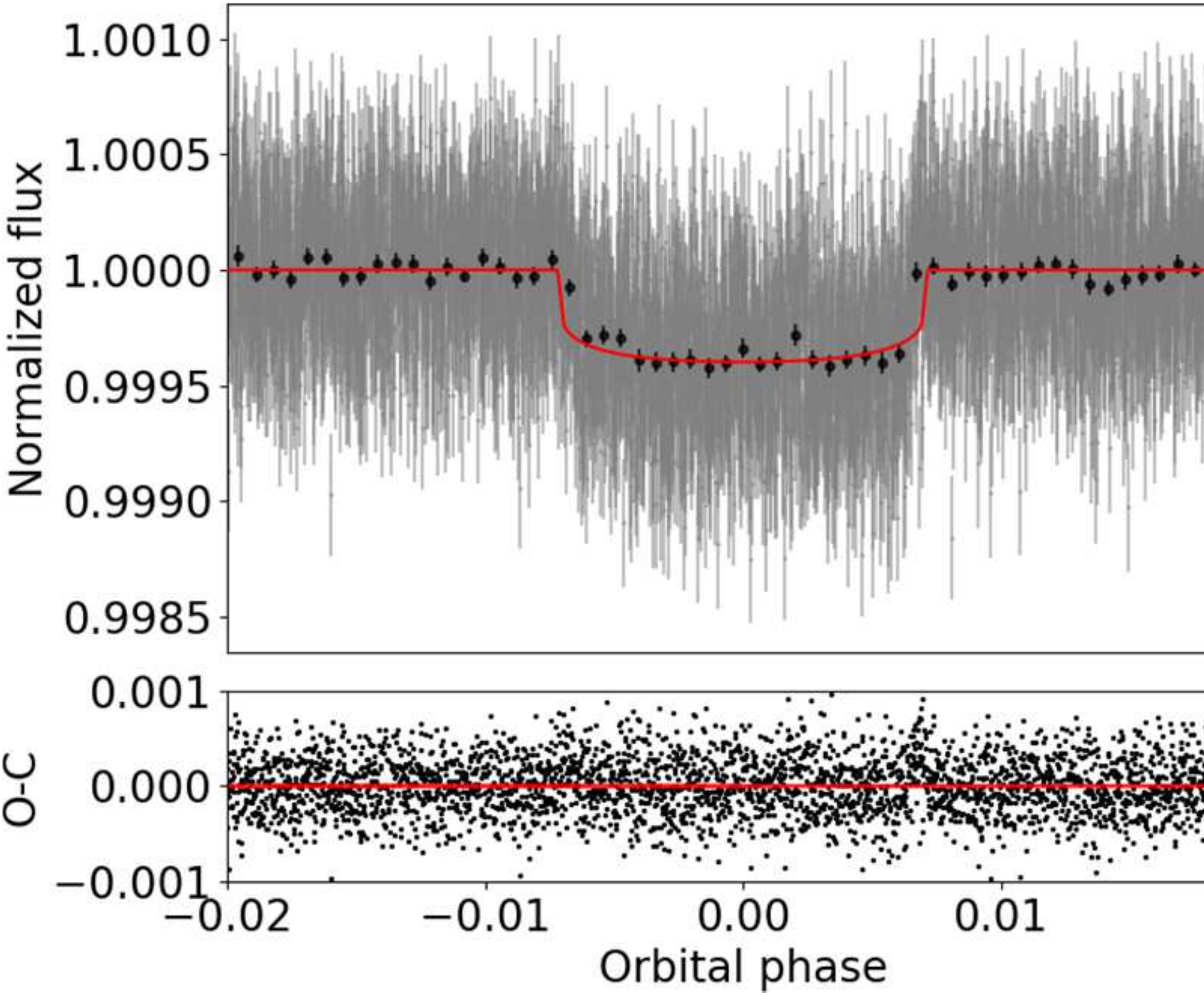}
    \caption{Top: The phase-folded transit light curve of HD 5278 b from 8 individual transit events. The best-fit transit model produced in the combined ESPRESSO+TESS data analysis in Sect. \ref{combinedESPRESSOTESS} is depicted by the red curve. The black dots represent data averaged over 13.8 min bins.} Bottom: the post-fit residuals.
    \label{fig:transfit}
\end{figure}

In order to see whether the amount of activity-induced RV variability captured in the jitter terms is expected for a star such as HD 5278, we cannot use a relation between the rms of the TESS light-curve and $v\sin i$ such as that in \citet{Mann2018}, as no variability in the photometry is detected (such a relation would in any case predict RV jitter $\sim10$ times larger than the measured one). If we use the  updated empirical relation between the \ion{Ca}{II} H \& K activity indicator and RV jitter presented by \citet{Hojjatpanah2020}, we would infer RV variability of $3.5\pm2.4$ m s$^{-1}$. This estimate is nominally higher than, but approximately within $1\sigma$ of, the measured values. Finally, the detailed simulations by \citet{Meunier2019} indicate that amplitudes of photometric variability of $\lesssim300$ ppm would be compatible with RV jitter around 1 m s$^{-1}$ for a star with the spectral type of HD 5278. Interestingly enough, given the typical value of $\log R^\prime_\mathrm{HK}$ we measure for the target, such low amplitudes of photometric modulation are expected, in a statistical sense, for quasi-pole-on configurations (see Figure 4 of \citealt{Meunier2019}).  

No significant orbital eccentricity is detected. We attempted orbital fits with either one or both companions on circular orbits, with no significant improvements in likelihood. Tidal circularization timescales for HD 5278 b are on the order of tens of Gyr (e.g., \citealt{Matsumura2008}), and larger still for the outer companion, indicating that there is no expectation for the orbits to be fully circularized. We therefore adopted the Keplerian solution as baseline, and report the corresponding $1\sigma$-level upper limits on $e$ in Table \ref{tab:planet_param}. 


\begin{table*}[ht!]
\centering
\caption{HD 5278 system parameters from separate analysis of TESS and ESPRESSO data. $1~\sigma$ uncertainties and upper limits are reported.}            
\begin{minipage}[t]{14.0cm} 
\setlength{\tabcolsep}{5.0mm}
\renewcommand{\footnoterule}{}                          
\begin{tabular}{l c c}        
\hline\hline
\noalign{\smallskip}
Parameter     &  Prior & Value \\
             \noalign{\smallskip}
             \hline
             \noalign{\smallskip}
             \noalign{\smallskip}
\multicolumn{3}{c}{\emph{Transit light curve parameters} } \\
\noalign{\smallskip}
Orbital period $P$ [days] & $\mathcal{U}$($-$inf,inf)& $14.33916_{-0.00004}^{+0.00005}$\\
\noalign{\smallskip}
Transit epoch $T_{ \rm c}\,\, [\rm BJD_{TDB}-2450000$] & $\mathcal{U}$($-$inf,inf)& $8680.6994_{-0.0007}^{+0.008}$ \\
\noalign{\smallskip}
Transit duration $T_{\rm 14}$ [days] &$\mathcal{U}$($-$inf,inf) & $0.1932_{-0.0014}^{+0.0015}$ \\
\noalign{\smallskip}
Radius ratio $R_{\rm p}/R_{\star}$ &$\mathcal{U}$($-$inf,inf) & $0.0188\pm0.0002$  \\
\noalign{\smallskip}
Inclination $i$ [deg] &$\mathcal{U}$(0.0,inf & $88.82_{-0.12}^{+0.13}$ \\
\noalign{\smallskip}
Jitter Sector 1 $\sigma_\mathrm{S1}$ &$\mathcal{U}$(0.0,inf) & $0.000051_{-0.000031}^{+0.000026}$ \\
\noalign{\smallskip}
Jitter Sector 12 $\sigma_\mathrm{S12}$ &$\mathcal{U}$($-$inf,inf) & $0.000129_{-0.000017}^{+0.000015}$ \\
\noalign{\smallskip}
Jitter Sector 13 $\sigma_\mathrm{S13}$ &$\mathcal{U}$($-$inf,inf)  & $0.000093_{-0.000022}^{+0.000018}$ \\
\noalign{\smallskip}
Jitter Sector 27 $\sigma_\mathrm{S27}$ &$\mathcal{U}$($-$inf,inf)  & $0.000169_{-0.000013}^{+0.000012}$ \\
\noalign{\smallskip}
$u_{1}$ \,\, [fixed] & & 0.24 \\
\noalign{\smallskip}
$u_{2}$ \,\, [fixed] & & 0.30 \\
\noalign{\smallskip}
\multicolumn{3}{c}{\emph{Spectroscopic orbit parameters} } \\
            \noalign{\smallskip}
$K_{\rm b}$ [m$\,s^{-1}$] & $\mathcal{U}$(0,5) & 1.96$^{+0.38}_{-0.35}$ \\
            \noalign{\smallskip}
$P_{\rm b}$ [days] & $\mathcal{N}$(14.3399,0.0032) & 14.33902$\pm$0.00007  \\ 
            \noalign{\smallskip}
$T_{c,\rm b}$ [BJD-2\,450\,000] & $\mathcal{N}$(8551.645659,0.002205) & 8551.645$\pm$0.001  \\
            \noalign{\smallskip}
$\sqrt{e_{\rm b}}\sin\omega_{\rm b}$ & $\mathcal{U}$(-1,1) & 0.354$^{+0.205}_{-0.325}$ \\
            \noalign{\smallskip}
$\sqrt{e_{\rm b}}\cos\omega_{\rm b}$ & $\mathcal{U}$(-1,1) &  $-0.217^{+0.212}_{-0.163}$ \\
            \noalign{\smallskip}
$K_{\rm c}$ [m$\,s^{-1}$] & $\mathcal{U}$(0,10) & 3.13$^{+0.35}_{-0.31}$ \\ 
            \noalign{\smallskip}
$P_{\rm c}$ [days] & $\mathcal{U}$(0,45) & 40.897$^{+0.173}_{-0.175}$  \\ 
            \noalign{\smallskip}
$T_{c,\rm c}$ [BJD-2\,450\,000] & $\mathcal{U}$(8420,8475) &  8431.8$\pm$1.3 \\ $\sqrt{e_{\rm c}}\sin\omega_{\rm c}$ & $\mathcal{U}$(-1,1) & $0.050^{+0.216}_{-0.230}$ \\
            \noalign{\smallskip}
$\sqrt{e_{\rm c}}\cos\omega_{\rm c}$ & $\mathcal{U}$(-1,1) &  $-0.010^{+0.222}_{-0.220}$ \\
            \noalign{\smallskip}
$\sigma_{\rm jit,1}$ [m$\,s^{-1}$] & $\mathcal{U}$(0,10) & 0.82$^{+0.42}_{-0.29}$  \\ 
            \noalign{\smallskip}
$\gamma_{\rm 1}$ [m$\,s^{-1}$] & $\mathcal{U}$(-30450,-30350) & $-30395.3\pm0.4$ \\
            \noalign{\smallskip}
$\sigma_{\rm jit,2}$ [m$\,s^{-1}$] & $\mathcal{U}$(0,10) & 1.49$^{+0.28}_{-0.30}$  \\ 
            \noalign{\smallskip}
$\gamma_{\rm 2}$ [m$\,s^{-1}$] & $\mathcal{U}$(-30450,-30350) & $-30396.6\pm0.3$ \\
            \noalign{\smallskip}

\noalign{\smallskip}
\hline
\noalign{\smallskip}
\multicolumn{3}{c}{\emph{Derived planetary parameters} } \\
\noalign{\smallskip}
Impact parameter $b_\mathrm{b}$ & & $0.43\pm0.05$ \\
            \noalign{\smallskip}
$e_{\rm b}$ &  & 0.22$^{+0.15}_{-0.13}$ ~($<0.29$)\\
            \noalign{\smallskip}
$\omega_{\rm b}$ ~[\rm rad] &  & 1.94$^{+0.57}_{-1.03}$ \\
            \noalign{\smallskip}
Orbital semi-major axis $a_{\rm b}$ [\rm au] & & 0.120$\pm$0.001\\
       \noalign{\smallskip}
Mass $M_{\rm b} ~[\rm M_\oplus]$ &  & $7.8_{-1.3}^{+1.4}$ \\
            \noalign{\smallskip}
Radius $R_{\rm b} ~[ \rm R_\oplus]$ &  & $2.45\pm0.05$ \\
            \noalign{\smallskip}
Density $\rho_{\rm b}$ [$\rm g\;cm^{-3}$] & & $2.9\pm0.5 $ \\
            \noalign{\smallskip}
Surface gravity log\,$g_{\rm b }$ [m s$^{-2}$] & & $12.6_{-2.3}^{+2,4}$ \\
            \noalign{\smallskip}
Equilibrium temperature $T_{\rm eq,b}$ [K] ~$^a$ &  & $943\pm13$\\
            \noalign{\smallskip}
Insolation $S_{\rm b}$ [\rm S$_\oplus$] &  & $132\pm5$\\
            \noalign{\smallskip}
$e_{\rm c}$ &  & 0.07$^{+0.08}_{-0.05}$ ~($<0.11$) \\
            \noalign{\smallskip}
$\omega_{\rm c}$ ~[\rm rad] &  & 0.59$^{+1.78}_{-2.42}$\\
            \noalign{\smallskip}
Minimum mass $M_{\rm c}\sin{i_{\rm c}}$ [\rm M$_\oplus$] & & 18.1$\pm$2.0\\
\noalign{\smallskip}
Orbital semi-major axis $a_{c}$ [au] & & 0.242$\pm$0.003\\
       \noalign{\smallskip}
\hline\hline       
\vspace{-0.5cm}
\footnotetext[1]{\scriptsize Black-body equilibrium temperature assuming a null Bond albedo and uniform 
heat redistribution to the night side.} \\
\end{tabular}
\end{minipage}
\label{tab:planet_param}  
\end{table*}


The physical companion identified in Gaia DR2 is located at a projected linear separation of 120 AU. Given the nominal masses for primary and secondary, this corresponds to a period in the neighborhood of 1100 years. It is then worthwhile investigating whether evidence of orbital motion from the companion is in fact already present in the high-precision ESPRESSO RVs of HD 5278. Following \citet{Torres1999}, given the expected companion mass, separation in the sky, and distance, the maximum expected RV slope is $\dot{\gamma}\simeq 1.6$ m s$^{-1}$ yr$^{-1}$. This is in principle already detectable in the ESPRESSO data. However, the residuals to the two-planet fit show no evidence for a long-term trend. We explored nevertheless this scenario adding a slope to the global model. We only derived a statistically non-significant $\dot{\gamma} = 1.48\pm0.67$ m s$^{-1}$ yr$^{-1}$, with very marginal Bayesian evidence in favour of the more complex model ($\ln\mathcal{Z}=-97.1$ and $\ln\mathcal{Z}=-96.2$ for the models without and with the slope, respectively). Things notwithstanding, we cannot claim a detection of the binary orbital motion. Only additional, long-term RV monitoring would allow for a clear statement to be made. Table \ref{tab:planet_param} reports the results of the model without the slope. 

\subsection{Combined TESS photometry + ESPRESSO RV analysis}\label{combinedESPRESSOTESS}


\begin{table*}[ht!]
\centering
\caption{HD 5278 system parameters derived from a joint ESPRESSO RVs+TESS photometry analysis. Errors and upper limits as in Table \ref{tab:planet_param}. }            
\begin{minipage}[t]{14.0cm} 
\setlength{\tabcolsep}{5.0mm}
\renewcommand{\footnoterule}{}                          
\begin{tabular}{l c c}        
\hline\hline
\noalign{\smallskip}
Parameter     &  Prior & Value \\
             \noalign{\smallskip}
             \hline
             \noalign{\smallskip}
             \noalign{\smallskip}
\multicolumn{3}{c}{\emph{Fitted system parameters} } \\
\noalign{\smallskip}
Radius ratio $R_{\rm p}/R_{\star}$ &$\mathcal{U}$(0.017,0.020) & $0.01874_{-00027}^{+0.00030}$  \\
\noalign{\smallskip}
Inclination $i$ [deg] &$\mathcal{U}$(85,90) & $89.27_{-0.48}^{+0.47}$ \\
\noalign{\smallskip}
Stellar density $\varrho_{\star}$ [$\varrho_\odot$] & $\mathcal{N}$(0.66,0.04) & $0.661_{-0.37}^{+0.039}$ \\
\noalign{\smallskip}
Jitter Sector 1 $\sigma_\mathrm{S1}$ &$\mathcal{U}$(0,0.003) & $0.000052_{-0.000031}^{+0.000025}$ \\
\noalign{\smallskip}
Jitter Sector 12 $\sigma_\mathrm{S12}$ &$\mathcal{U}$(0,0.003) & $0.000129_{-0.000016}^{+0.000015}$ \\
\noalign{\smallskip}
Jitter Sector 13 $\sigma_\mathrm{S13}$ &$\mathcal{U}$(0,0.003)  & $0.000094_{-0.000022}^{+0.000019}$ \\
\noalign{\smallskip}
Jitter Sector 27 $\sigma_\mathrm{S27}$ &$\mathcal{U}$(0,0.003)  & $0.000169_{-0.000013}^{+0.000012}$ \\
\noalign{\smallskip}
$u_{1}$ \,\, [fixed] & & 0.24 \\
\noalign{\smallskip}
$u_{2}$ \,\, [fixed] & & 0.30 \\
\noalign{\smallskip}
$K_{\rm b}$ [m$\,s^{-1}$] & $\mathcal{U}$(0,5) & 1.93$^{+0.34}_{-0.35}$ \\
            \noalign{\smallskip}
$P_{\rm b}$ [days] & $\mathcal{U}$(14,14.5) & 14.339156$_{-0.000047}^{+0.000049}$  \\ 
            \noalign{\smallskip}
$T_{c,\rm b}$ [BJD-2\,450\,000] & $\mathcal{U}$(8681,8682) & 8680.69932$_{-0.00081}^{+0.00083}$  \\
            \noalign{\smallskip}
$\sqrt{e_{\rm b}}\sin\omega_{\rm b}$ & $\mathcal{U}$(-1,1) & 0.09$^{+0.09}_{-0.12}$ \\
            \noalign{\smallskip}
$\sqrt{e_{\rm b}}\cos\omega_{\rm b}$ & $\mathcal{U}$(-1,1) &  $-0.25^{+0.24}_{-0.15}$ \\
            \noalign{\smallskip}
$K_{\rm c}$ [m$\,s^{-1}$] & $\mathcal{U}$(0,6) & 3.17$_{-0.32}^{+0.34}$ \\ 
            \noalign{\smallskip}
$P_{\rm c}$ [days] & $\mathcal{U}$(0,45) & 40.87$^{+0.18}_{-0.17}$  \\ 
            \noalign{\smallskip}
$T_{c,\rm c}$ [BJD-2\,450\,000] & $\mathcal{U}$(8420,8475) &  8431.8$^{+1.2}_{-1.4}$ \\ 
            \noalign{\smallskip}
$\sqrt{e_{\rm c}}\sin\omega_{\rm c}$ & $\mathcal{U}$(-1,1) & $0.01^{+0.20}_{-0.21}$ \\
            \noalign{\smallskip}
$\sqrt{e_{\rm c}}\cos\omega_{\rm c}$ & $\mathcal{U}$(-1,1) &  $-0.006^{+0.227}_{-0.217}$ \\
            \noalign{\smallskip}
$\sigma_{\rm jit,1}$ [m$\,s^{-1}$] & $\mathcal{U}$(0,10) & 0.83$^{+0.47}_{-0.48}$  \\ 
            \noalign{\smallskip}
$\gamma_{\rm 1}$ [m$\,s^{-1}$] & $\mathcal{U}$(-30450,-30350) & $-30395.4\pm0.4$ \\
            \noalign{\smallskip}
$\sigma_{\rm jit,2}$ [m$\,s^{-1}$] & $\mathcal{U}$(0,10) & $1.51^{+0.25}_{-0.20}$  \\ 
            \noalign{\smallskip}
$\gamma_{\rm 2}$ [m$\,s^{-1}$] & $\mathcal{U}$(-30450,-30350) & $-30396.4\pm0.3$ \\
            \noalign{\smallskip}

\noalign{\smallskip}
\hline
\noalign{\smallskip}
\multicolumn{3}{c}{\emph{Derived planetary parameters} } \\
\noalign{\smallskip}
Impact parameter $b_\mathrm{b}$ & & $0.27_{-0.18}^{+0.17}$ \\
	    \noalign{\smallskip}
$a/R_{\star}$ & & $22.4^{+0.5}_{-0.9}$ \\
	    \noalign{\smallskip}
Transit duration $T_{\rm 14}$ [days] & & $0.200_{-0.012}^{+0.010}$ \\
            \noalign{\smallskip}
$e_{\rm b}$ &  & 0.08$^{+0.09}_{-0.05}$ ~($<0.12$)\\
            \noalign{\smallskip}
$\omega_{\rm b}$ ~[\rm rad] &  & 2.36$^{+0.59}_{-4.88}$ \\
            \noalign{\smallskip}
Orbital semi-major axis $a_{\rm b}$ [\rm au] & & 0.1202$\pm$0.0013\\
       \noalign{\smallskip}
Mass $M_{\rm b} ~[\rm M_\oplus]$ &  & $7.8_{-1.4}^{+1.5}$ \\
            \noalign{\smallskip}
Radius $R_{\rm b} ~[ \rm R_\oplus]$ &  & $2.45\pm0.05$ \\
            \noalign{\smallskip}
Density $\rho_{\rm b}$ [$\rm g\;cm^{-3}$] & & $2.9_{-0.5}^{+0.6} $ \\
            \noalign{\smallskip}
Surface gravity log\,$g_{\rm b }$ [m s$^{-2}$] & & $12.8_{-2.3}^{+2.4}$ \\
            \noalign{\smallskip}
Equilibrium temperature $T_{\rm eq,b}$ [K] ~$^a$ &  & $943\pm13$\\
            \noalign{\smallskip}
Insolation $S_{\rm b}$ [\rm S$_\oplus$] &  & $132\pm5$\\
            \noalign{\smallskip}
$e_{\rm c}$ &  & 0.07$^{+0.08}_{-0.04}$ ~($<0.10$) \\
            \noalign{\smallskip}
$\omega_{\rm c}$ ~[\rm rad] &  & 0.14$^{+2.03}_{-2.14}$\\
            \noalign{\smallskip}
Minimum mass $M_{\rm c}\sin{i_{\rm c}}$ [\rm M$_\oplus$] & & 18.4$_{-1.9}^{+1.8}$\\
\noalign{\smallskip}
Orbital semi-major axis $a_{c}$ [au] & & 0.2416$\pm$0.0027\\
       \noalign{\smallskip}
\hline\hline       
\vspace{-0.5cm}
\end{tabular}
\end{minipage}
\label{tab:planet_param_combined}  
\end{table*}

\begin{figure}[t!]
    \centering
    \centering
    \includegraphics[width=0.45\textwidth, angle=0]{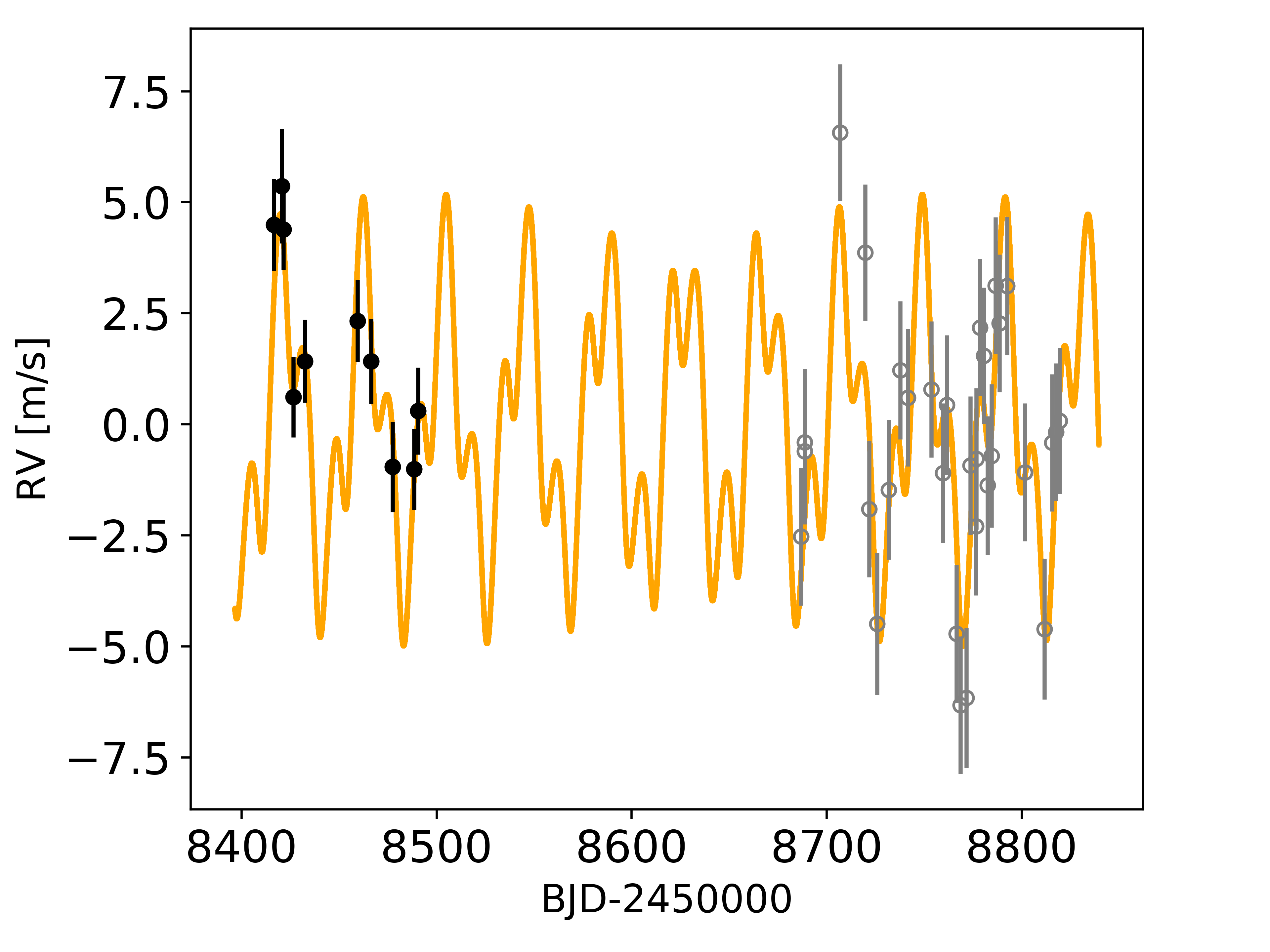} \\
    \includegraphics[width=0.45\textwidth, angle=0]{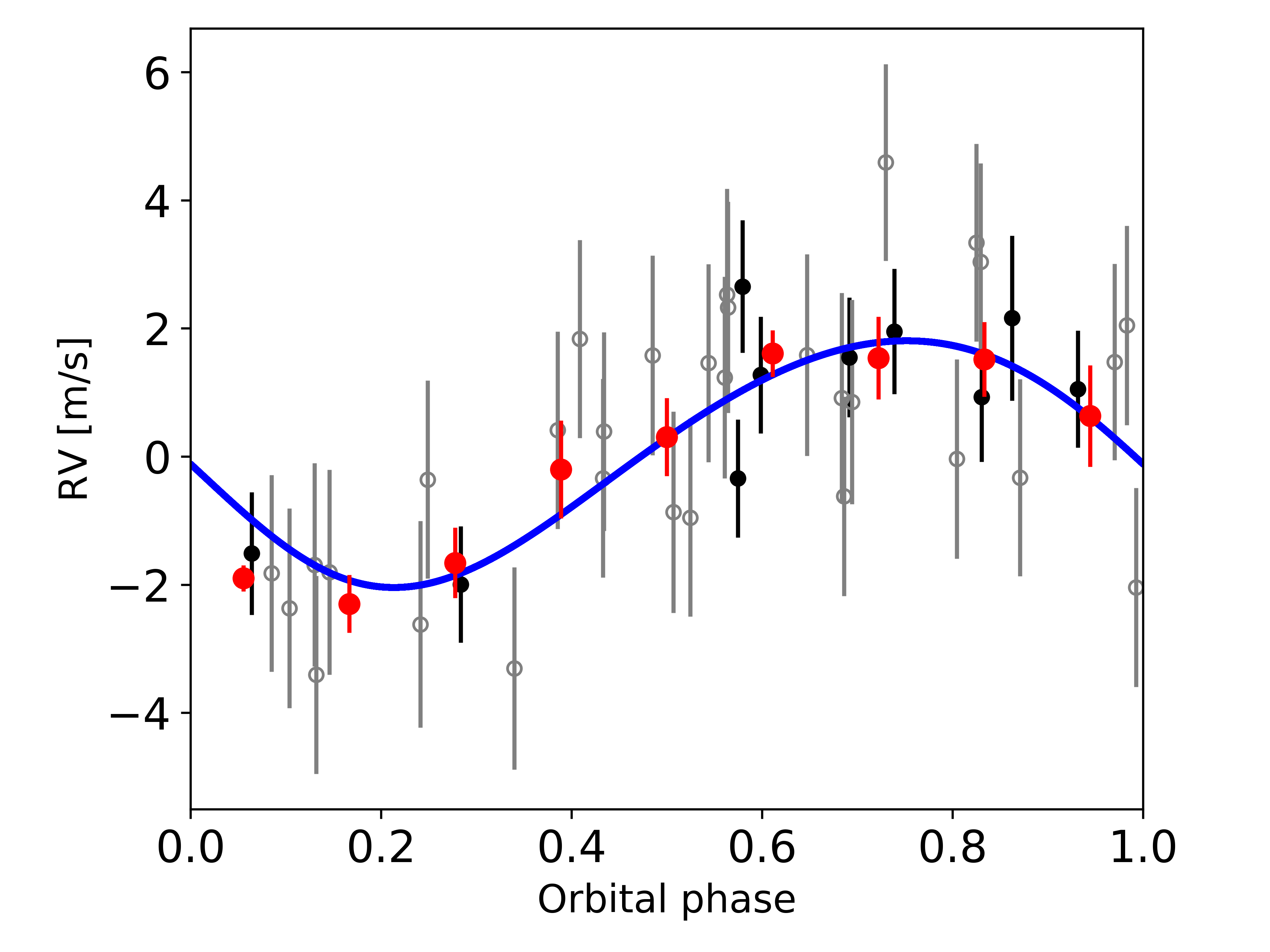} \\
    \includegraphics[width=0.45\textwidth, angle=0]{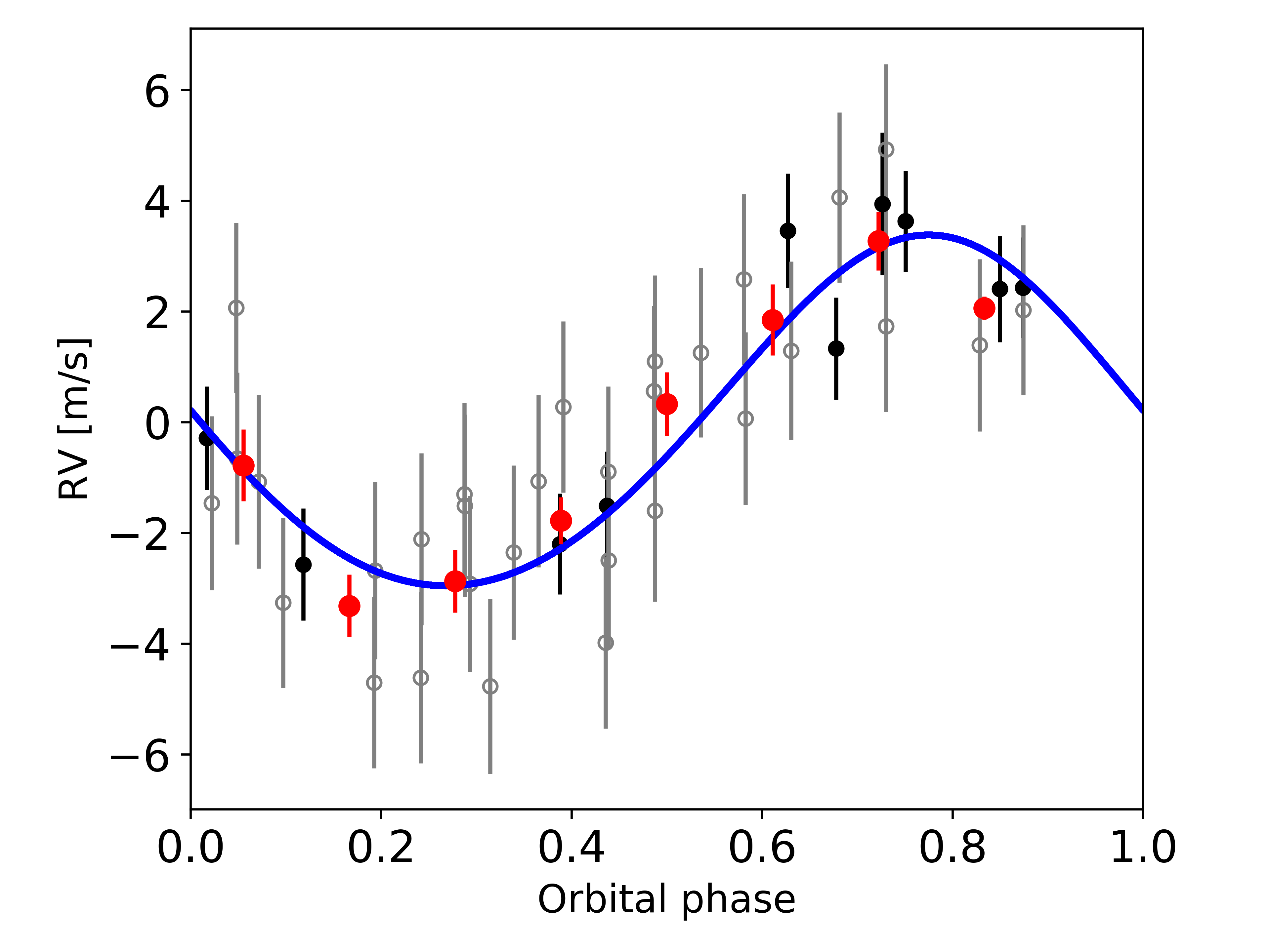} \\
    \caption{Top: the ESPRESSO RV times series for HD 5278 with overplotted the best-fit two-planet solution (yellow curve) obtained in the joint ESPRESSO RV + TESS photometry analysis. Black filled and grey open circles correspond to ESPRESSO observations obtained before and after the technical intervention. Center: the phase-folded best-fit orbital model for the transiting companion HD 5278 b (blue curve). Individual observations are colour-coded as in the top panel, while the large red filled dots show them binned in phase. Bottom: the same for the non-transiting companion HD 5278 c. In all panels, formal uncertainties have been inflated adding in quadrature the RV jitter values reported in Table \ref{tab:planet_param_combined}.}
    \label{fig:orbfit}
\end{figure}

\begin{figure}
    \centering
    \includegraphics[width=0.49\textwidth, angle=0]{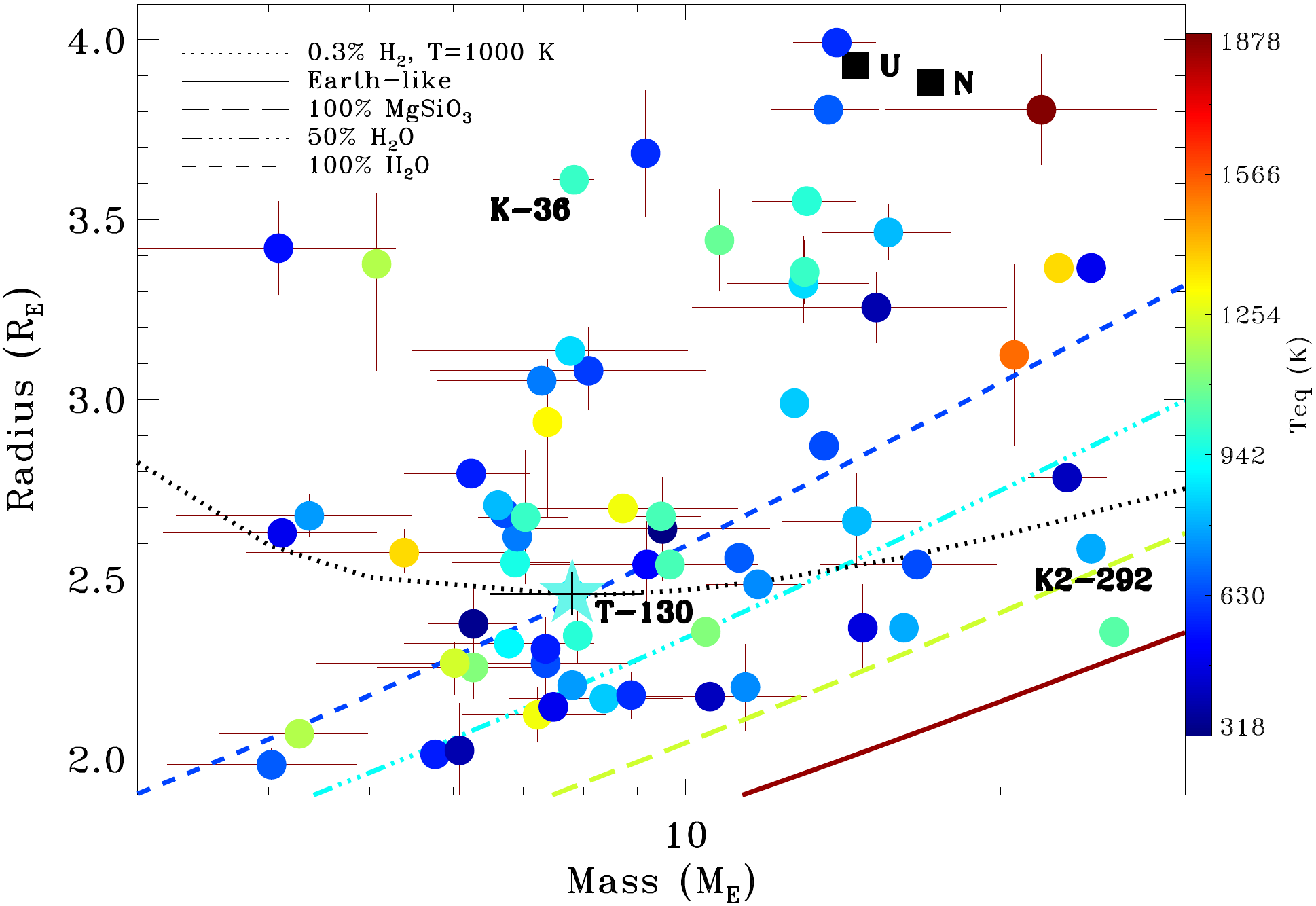}
\caption{Mass-radius diagram of sub-Neptunes with masses detected at the $3\sigma$ level (or better), including HD 5278 b (star). The objects are color-coded by their equilibrium temperature. The different curves depict internal structure models of variable composition from \citet{Zeng2013} and \citet{Zeng2019} (as reported in the legend): Earth-like (32.5\% Fe + 67.5\% MgSiO$_3$), pure rock (100\% MgSiO$_3$), 50\% Earth-like rocky core + 50\% H$_2$O layer by mass, 100\% H$_2$O mass, Earth-like rocky core (99.7\%) + 0.3\% H$_2$ envelope by mass at a temperature of 1000 K). The location of Uranus and Neptune is also indicated. Data of transiting systems used for this and the following plots are taken from the TEPCat catalogue as of September 2020. Along with TOI-130 b (T-130 for simplicity), Kepler-36 c and K2-292 b (K-36 and K2-292 for simplicity), which are discussed in the text, are also indicated in the plot.}
    \label{mrrelation}
\end{figure}

While the system parameters derived based on the separate analysis of the TESS light curve and ESPRESSO RVs are robust, it is desirable to perform a joint analysis of the combined datasets for two reasons: 1) it is possible to place tighter constraints on orbit eccentricity and derive self-consistent uncertainties in the model parameters taking into account correlations; 2) given the evidence for the second companion in the RV data, it is possible to derive self-consistent ephemeris and verify that HD 5278 c did not undergo an inferior conjunction during the timespan of the TESS observations.

The joint analysis of the TESS light curve and ESPRESSO RV time-series was carried out with two different Bayesian techniques: i) \texttt{MultiNestv3.10}, through its \texttt{pyMultiNest} wrapper, coupled to the code \texttt{batman} \citep{batman05}, which was setup to run with 500 live points; ii) a DE-MCMC method as for the transit fitting, following the same implementation as in \citet{Bonomo2014,Bonomo2015}. The model parameters in the two approaches are the same, with the only exception that the former method fits for the transit stellar density (constrained by the value reported in Table \ref{tab:starprop}), and the latter for the transit duration.

As a preliminary step, We first ran the \texttt{MultiNest}-based algorithm adopting a 1-planet model (only the transiting planet). This allowed us to perform model selection in a straightforward manner by comparing the resulting $\ln\mathcal{Z}$ in the joint analysis with that obtained in the case of the 2-planet model. The corresponding natural logarithm of evidence ratio is $\Delta\ln\mathcal{Z} = \ln\mathcal{Z}(\mathrm{2 planet}) - \ln\mathcal{Z}(\mathrm{1 planet}) = 29337.6 - 29324.8 = +12.8$. We can therefore conclude that there is, as expected, strong evidence in favour of the 2-planet model (e.g., \citealt{Nelson2020}). 

For the two-planet model the parameter best values and uncertainties as determined by both techniques are fully consistent and, for this reason, we report in Table \ref{tab:planet_param_combined} the results of the analysis from the first technique only. Figure \ref{fig:transfit} shows the phase-folded transit light-curve along with the best-fit transit model. Figure \ref{fig:orbfit} shows the two-planet fit superposed to the time-series of the ESPRESSO RVs and the phase-folded plots of the best-fit  orbits for both planets (with error bars taking into account the fitted values of RV jitter). Finally, the corner plots of Figures \ref{fig:posteriors_phot} and \ref{fig:posteriors_spec} summarize the results of the posterior distributions of the model parameters. 

Overall, the results are in excellent agreement with those derived with the individual analysis of ESPRESSO RVs and TESS photometry. The transit of HD 5278 b is found to be slightly more central than in the case of the separate analysis of the TESS light curve. Tighter constraints are placed on the eccentricity of HD 5278 b, which is now found to be $e_b<0.12$. Based on the updated ephemeris derived for HD 5278 c, if transiting the planet should have been spotted by TESS once in each of the four Sectors analyzed. We carefully inspected the TESS light curve, found no evidence of such events, and therefore conclude that HD 5278 c indeed does not transit. We elect to adopt the numbers in Table \ref{tab:planet_param_combined} as our fiducial system parameters.

\section{Discussion}

\subsection{The sub-Neptune HD 5278 b}\label{hd5278b_comp}

The RV semi-amplitude of the transiting companion ($K_\mathrm{b} = 1.93^{+0.34}_{-0.35}$ \ms) is clearly detected with ESPRESSO at $>5\sigma$ level. Given the stellar properties derived in Sect. \ref{starprop}, we infer $M_{\rm b} = 7.8_{-1.4}^{+1.5}$ M$_\oplus$, $R_{\rm b} = 2.45\pm0.05$ R$_\oplus$, and a planetary mean density $\rho_{\rm b} = 2.9_{-0.5}^{+0.6}$ g cm$^{-3}$. In the mass-radius diagram (Figure \ref{mrrelation}) of sub-Neptunes with $\geq3\sigma$ mass determinations it occupies a seemingly unremarkable location. Its density is not particularly low, compared to an object with similar mass but much larger radius such as Kepler-36 c \citep{Carter2012}. Its density is not particularly high either, compared to an object with almost identical radius but much larger mass such as K2-292 b \citep{Luque2019}. Its size makes HD 5278 b sit atop the peak of the radius distribution on the outer side of the radius valley \citep{Fulton2017,Fulton2018}. 

As shown in Figure \ref{mrrelation}, HD 5278 b's composition could be described as corresponding to a $\sim100\%$ water world with a minimal $H_2$-rich atmosphere \citep{Zeng2019}, or a massive rocky planet surrounded by a  H$_2$-dominated envelope of $<1\%$ in planet mass \citep{Lopez2014,Zeng2019}. The planet's mass and radius also fit the definition of an ice-dominated planet with a thin or non-existent H$_2$/He atmosphere \cite{Dorn2017,Zeng2019,Venturini2020}, or that of a water-rich world with a significant fraction (20-50\%) of water in super-critical steam state, and thus with an H$_2$O-dominated atmosphere and a more or less relevant H$_2$/He envelope \citep{Madhusudhan2020,Turbet2020,Mousis2020,Kite2020}. 

Using the measured mass, radius and density of HD 5278 b, we performed our own internal structure retrieval calculations (see \citealt{Mortier2020}), assuming the Fe, Si, and Mg stellar abundance ratios from Table \ref{tab:starprop} as proxies for the disk composition at the time of planet formation, and with the constraint that the planetary bulk abundance matches the stellar one. We explored two cases: a) a fully differentiated planet that includes a water layer, and in which the gas envelope is solely composed of H and He; b) a 'dry' model, where the planet interior structure is composed of only an iron core and a silicate mantle. In the first case, we find core, mantle, and water mass fractions of $0.13^{+0.08}_{-0.09}$, $0.56^{+0.15}_{-0.14}$, and $0.30^{+0.15}_{-0.17}$, respectively, where the error bars correspond to the 5\% and 95\% quantiles (approximately $2\sigma$). The gas mass is small (upper 95 \% quantile equal to $\sim 0.03 $ M$_\oplus$), and the envelope contributes approximately 17\% of the observed radius ($0.42^{+0.17}_{-0.16}$ R$_\oplus$). In the second case, we obtain $0.19^{+0.11}_{-0.12}$ and  $0.81^{+0.12}_{-0.11}$ for the core and mantle mass fractions, respectively. The gas envelope corresponds to around 4\% of the planet mass, and contributes 30\% of the observed planet radius ($0.74^{+0.05}_{-0.05}$ R$_\oplus$). 

The corner plots with the posterior distributions of the most relevant parameters adjusted in our internal structure retrieval analysis are shown in Figures \ref{fig:posteriors_model1} and \ref{fig:posteriors_model2}. As expected, they show how rather different compositions exactly reproduce the observed mass and radius of HD 5278 b. However, from a theoretical viewpoint, sub-Neptunes are primarily expected to have formed beyond the water ice line (e.g. \citealt{Raymond2018,Bitsch2019,Venturini2020}). Therefore, model a) is expected to be a closer representation of the composition of a planet in the second peak of the radius valley, such as HD 5278 b. We get back to this point in Sect. \ref{physprop}. 

\begin{figure}
    \centering
    \includegraphics[width=0.49\textwidth, angle=0]{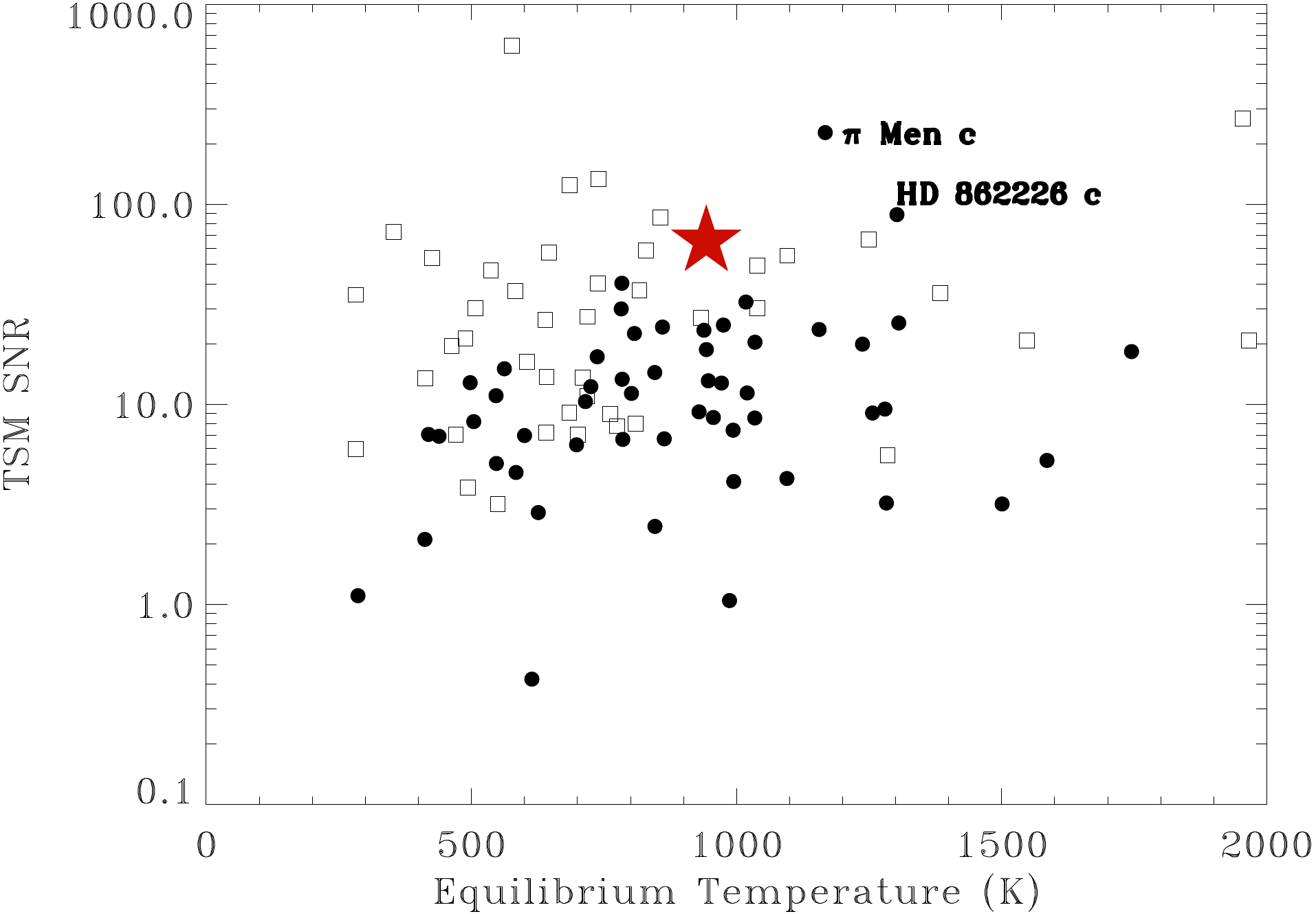}
    \caption{S/N for transmission spectroscopy observations with JWST (TSM; \citealt{Kempton2018}) vs. equilibrium temperature for sub-Neptunes with a measured mass, including HD 5278 b (red star). Black filled dots and empty squares identify the sample of planets around stars with $T_\mathrm{eff}<5400$ K and $T_\mathrm{eff}>5400$ K, respectively.}
    \label{rankmetrics}
\end{figure}

As mentioned in the Introduction, breaking the degeneracy between core mass and atmospheric metallicity for sub-Neptunes can only be attained via atmospheric characterization measurements (particularly in the case of mostly cloud-free atmospheres). In this respect, HD 5278 b is a very interesting target. For the case of low resolution spectroscopy in space (with e.g. JWST), using simple metrics for estimating the S/N in transmission and emission \citep{Gillon2016,Niraula2017,Kempton2018}, the planet always ranks highest amongst sub-Neptunes with $T_\mathrm{eq} < 1000$ K orbiting G-F dwarfs ($T_\mathrm{eff}>5400$ K). When all regimes of equilibrium temperature are considered, HD 5278 b ranks third, after HD 39091/$\pi$ Mensae c \citep{Huang2018} and HD 86226 c \citep{Teske2020}. HD 5278 b is high up in the ranking even when objects around cooler primaries are included (see Figure \ref{rankmetrics} for an example). For the case of F-G-type primaries, HD 5278 is more than 2 mag fainter than HD 39091 at $K$-band, somewhat alleviating the practical difficulties in achieving photon-limited observations of very bright stars with JWST. Even in the case of very efficient production of photochemical hazes, HD 5278 b belongs to a class of sub-Neptunes (with $T_\mathrm{eq} < 1000$ K ) that might show detectable molecular features via transmission spectroscopy with JWST \cite{Kawashima2019}.

HST observations of a few sub-Neptunes have returned featureless transmission spectra (e.g., \citealt{Kreidberg2014,Knutson2014}), and although there are successful detections of molecular absorption reported in the most recent literature for such objects (e.g., \citealt{Tsiaras2019,Benneke2019,Guilluy2020,Kreidberg2020}), this can naturally be seen as a potential limitation. However, the sample size is still very small, and new characterization measurements are mandatory to advance the field. HD 5278 b is an appealing target also for HST, orbiting a late F-type primary (a mostly unexplored spectral type) that is apparently showing very low levels of stellar activity (thus simplifying procedures to calibrate out such contaminating effects in transmission spectra). Finally, with $T_\mathrm{eq}\lesssim 1000$ K, the planet's atmosphere might be at the boundary between the transition regime from CH$_4$ to CO gas as the dominant carrier of carbon \citep{Moses2013,Hu2014} and the high-temperature range where relatively few cloud species are expected to condense and organic hazes are unstable, so that atmospheres are more likely to have large spectral features (e.g., \citealt{Fraine2014,Wakeford2017}). HD 5278 b is therefore less likely to show the presence of clouds and hazes. 

In principle, evidence of ongoing atmospheric escape in HD 5278 b's could be searched for probing the planet's upper atmosphere at high spectral resolution in transmission. The observing channel of choice would be the 10,833 \AA\, helium triplet feature, a better suited diagnostic than the hydrogen Lyman $\alpha$ line in the UV, given the distance of the target\footnote{At $\sim60$ pc, HD 5278 sits approximately at the distance limit for useful measurements of the stellar Lyman $\alpha$ line flux, due to its exponential decay caused by absorption by neutral hydrogen atoms in the interstellar medium.}. However, while the escaping/extended atmospheres of several hot Jupiters have successfully been probed in this way (e.g., \citealt{Allart2018,Spake2018,Nortmann2018,Guilluy2020}), recent attempts at detecting the upper atmospheres in sub-Neptunes using the helium infrared triplet have shown how both observational and theoretical challenges must first be met \cite{Kasper2020}.

If the low levels of photometric variability of the primary are indeed indicative of a likely quasi-pole-on configuration (see previous discussion on the \citealt{Meunier2019} simulation results), a finding corroborated by the low value of $v\sin i$ for a star with the temperature of HD 5278 (pointing to a possibly large obliquity), this raises the possibility of a significant spin-orbit misalignment for HD 5278 b. Given the planetary and stellar radius and v$\sin i$ value, the expected amplitude of the Rossiter-McLaughlin effect (for a planetary orbital plane well aligned with the stellar spin axis) is on the order of 1.5 m s$^{-1}$, which appears within reach with ESPRESSO (aiming at a full RV sequence during a transit window), given the brightness of the host. 

\subsection{The non-transiting Neptune HD 5278 c and system evolution}

We have shown how Bayesian evidence strongly favours the two-planet model. This, alongside the lack of activity-induced RV signals, helped us build a solid case for explaining the RV modulation with a 41-day period seen in the ESPRESSO data in terms of an outer, non-transiting companion. The $10\sigma$-level detection ($K_{\rm c} = 3.17^{+0.34}_{-0.32}$ \ms) of its RV semi-amplitude implies a minimum mass $M_{\rm c}\sin{i_{\rm c}} = 18.4_{-1.9}^{+1.8}$ M$_\oplus$, making it a Neptune-mass planet. We found no evidence of its transit in the TESS photometry, and therefore we can infer an upper limit to its orbital inclination ($i_c < \arccos(a/R_\star)^{-1}$) $i_c < 88.68$ deg. This in turn implies a lower limit to the true mass of HD 5278 c M$_\mathrm{p,c} > 18.401$ M$_\oplus$. 

The HD 5278 two-planet system is Hill-stable, following standard analytical criteria (e.g., \citealt{Giuppone2013}), even assuming the eccentricities of the two companions match the upper limits derived in Table \ref{tab:planet_param}. Stability is violated only in the limit of a true mass of HD 5278 c in excess of 200 M$_\oplus$, which would imply a true inclination angle of $i_{\rm c}\lesssim 5$ deg. Taken at face value, the non-zero eccentricity of HD 5278 b and the hint of its possibly misaligned orbit could indicate a relatively 'active' evolutionary history for the HD 5278 system, in which dynamical instabilities and chaos have played an important role (e.g., \citealt{Chambers2001,Laskar2017,Turrini2020}, and references therein), a scenario that is typically favored for low multiplicity systems such as the one under investigation here. However, precisely because of the lack of a statistically significant determination of non-zero orbital eccentricity for either of the two planets (and in absence of an actual measurement of the spin-orbit angle), the present orbital configuration of the HD 5278 system is also in line with the findings of \citet{Mills2019}, who showed that multiples in the Kepler field have typically $e< 0.1$. In this case, the evolution history of the system is expected to have been less influenced by chaos and instabilities.

The Kepler mission multi-planet sample has been shown to have a distribution of period ratios with excesses of systems (“peaks”) just wide of first-order (3:2, 2:1) and second-order (5:3, 3:1) mean motion resonances (MMR), and corresponding deficits (“troughs”) just short of these commensurabilities (e.g., \citealt{Lissauer2011,Fabrycky2014,Delisle2014,Xu2017,Choksi2020}). The HD 5278 system is unremarkable in this sense, with a period ratio deviation from the closest second-order MMR (3:1) of -0.148, typically a factor of 10 larger than those corresponding to the peaks and troughs of near-resonance configurations identified in the original Kepler sample of multiples. The formation of a low-mass system such as the one we have uncovered around HD 5278 is still open for debate. As the two planets do not lie near a period commensurability and they are not piled up at very short periods, formation (mostly) in-situ in gas-poor conditions is viable (e.g., \citealt{Lithwick2012,Ogihara2018,Terquem2019,Choksi2020}, and references therein). Alternatively, formation in the gas-rich regions of the protoplanetary disk followed by more or less 'clean' disc-driven migration and resonance capture can also possibly describe the present architecture of the HD 5278 system, provided the two planets subsequently escaped resonance due to dynamical instability effects driven by a variety of other physical processes (e.g., \citealt{Goldreich2014,Izidoro2017,Lambrechts2019,Raymond2020}, and references therein). 

Recent studies \citep{Bryan2019} have shown that planetary systems with inner super Earths and sub-Neptunes are often accompanied by outer companions of unclear nature (be they massive planets or brown dwarfs). If confirmed by future observations, the fact that the marginal RV trend we measured in the ESPRESSO data might be compatible with the gravitational influence of the distant binary companion to HD 5278 would rule out the existence of massive substellar objects at shorter separations. In the \citet{Kervella2019} catalog of proper motion anomalies the sensitivity curve for HD 5278 implies that companions with masses of 0.37 M$_J$ au$^{-1/2}$ are ruled out based on the combination of Gaia and Hipparcos astrometry alone. Based on the analytical formulation in \citet{Kervella2019}, at the exact separation of HD 5278 c the detectable mass from proper motion anomaly is found to be around 60 M$_\mathrm{J}$, placing uninteresting constraints on the true mass of the outer planet in the system. Taking into account the loss of efficiency in companion mass determination for orbital periods vastly exceeding the Hipparcos-Gaia time baseline, this technique would be marginally sensitive to the low-mass binary companion at $\sim120$ au. Massive ($\gtrsim 10$ M$_J$) companions at $\sim10$ au would however be clearly detectable, and the lack of a statistically significant Hipparcos-Gaia proper motion difference qualitatively points towards their absence, in agreement with the evidence from the ESPRESSO RVs. 

\subsection{On the physical properties of sub-Neptunes} \label{physprop}

\begin{figure*}
\begin{tabular}{c c}
    \centering
    \includegraphics[width=0.49\textwidth, angle=0]{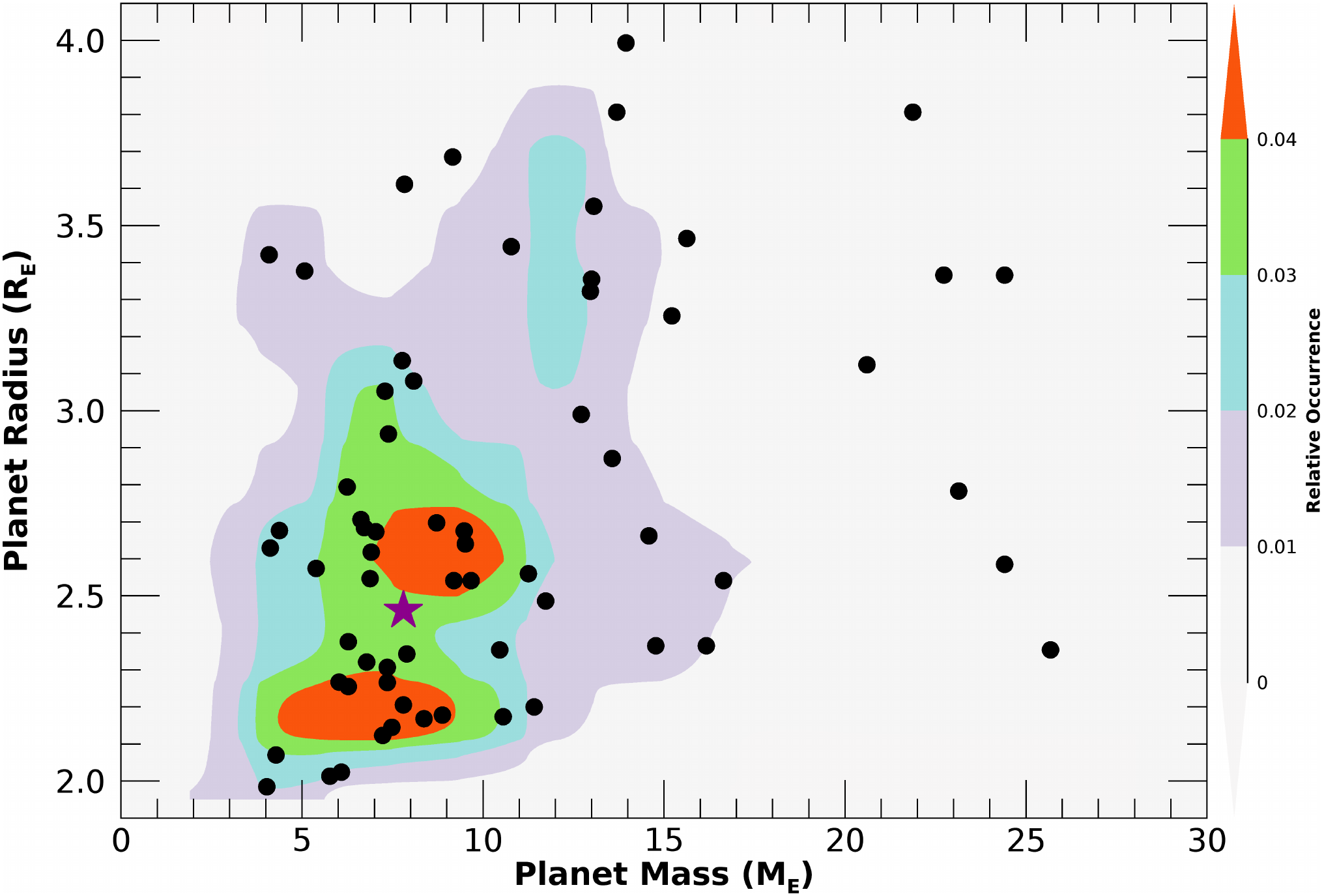} &
    \includegraphics[width=0.49\textwidth, angle=0]{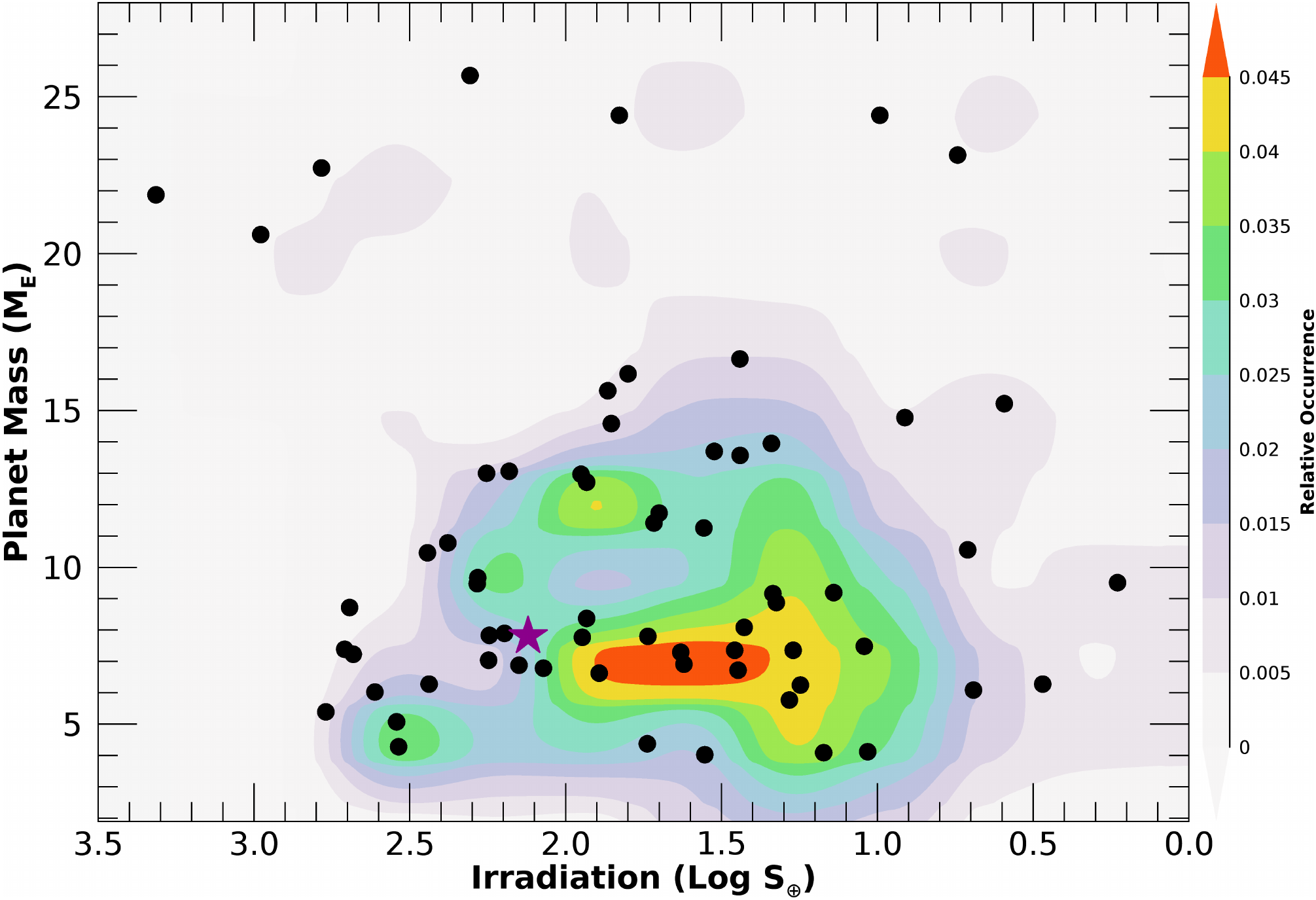} \\
    \includegraphics[width=0.49\textwidth, angle=0]{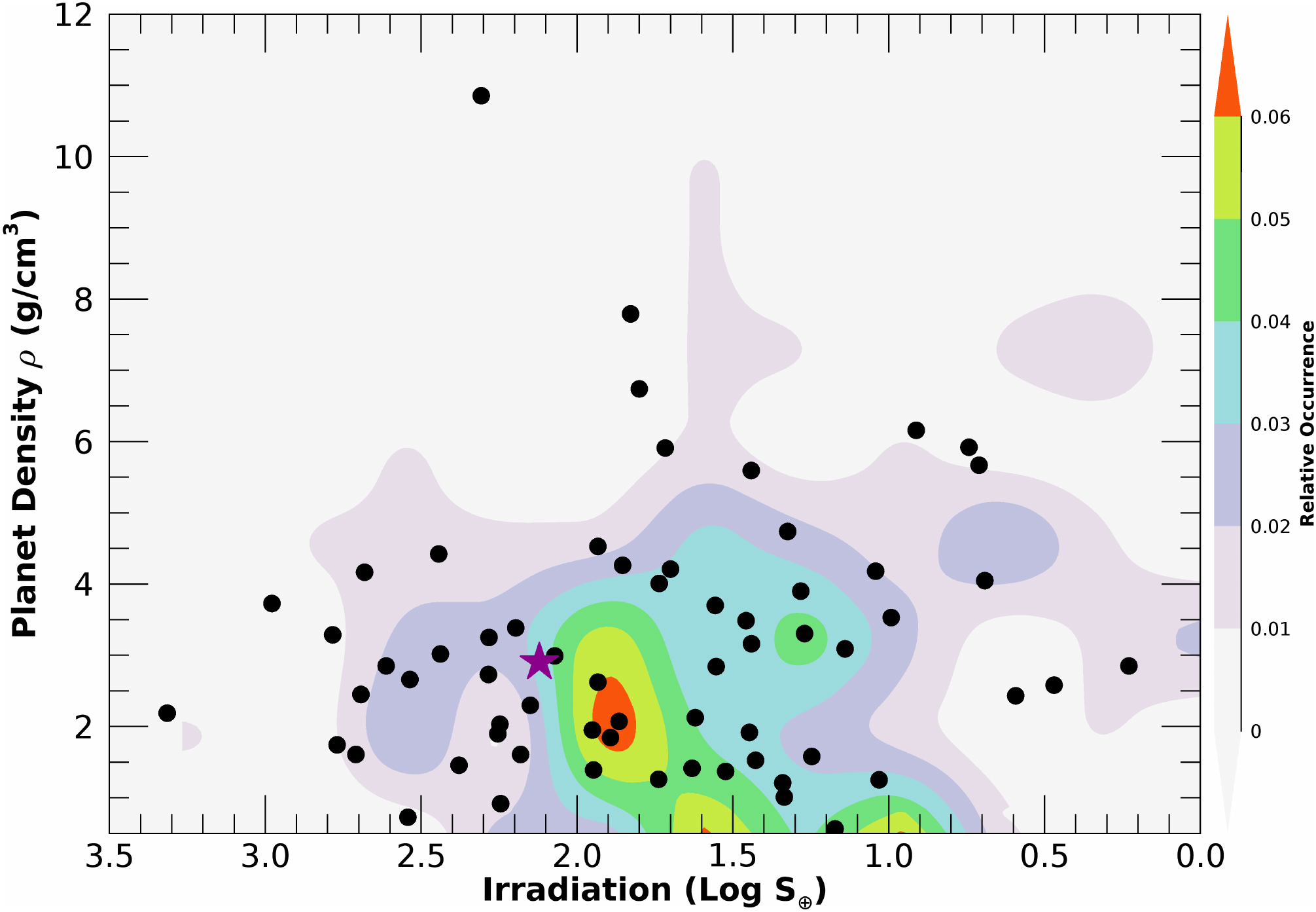} &
    \includegraphics[width=0.49\textwidth, angle=0]{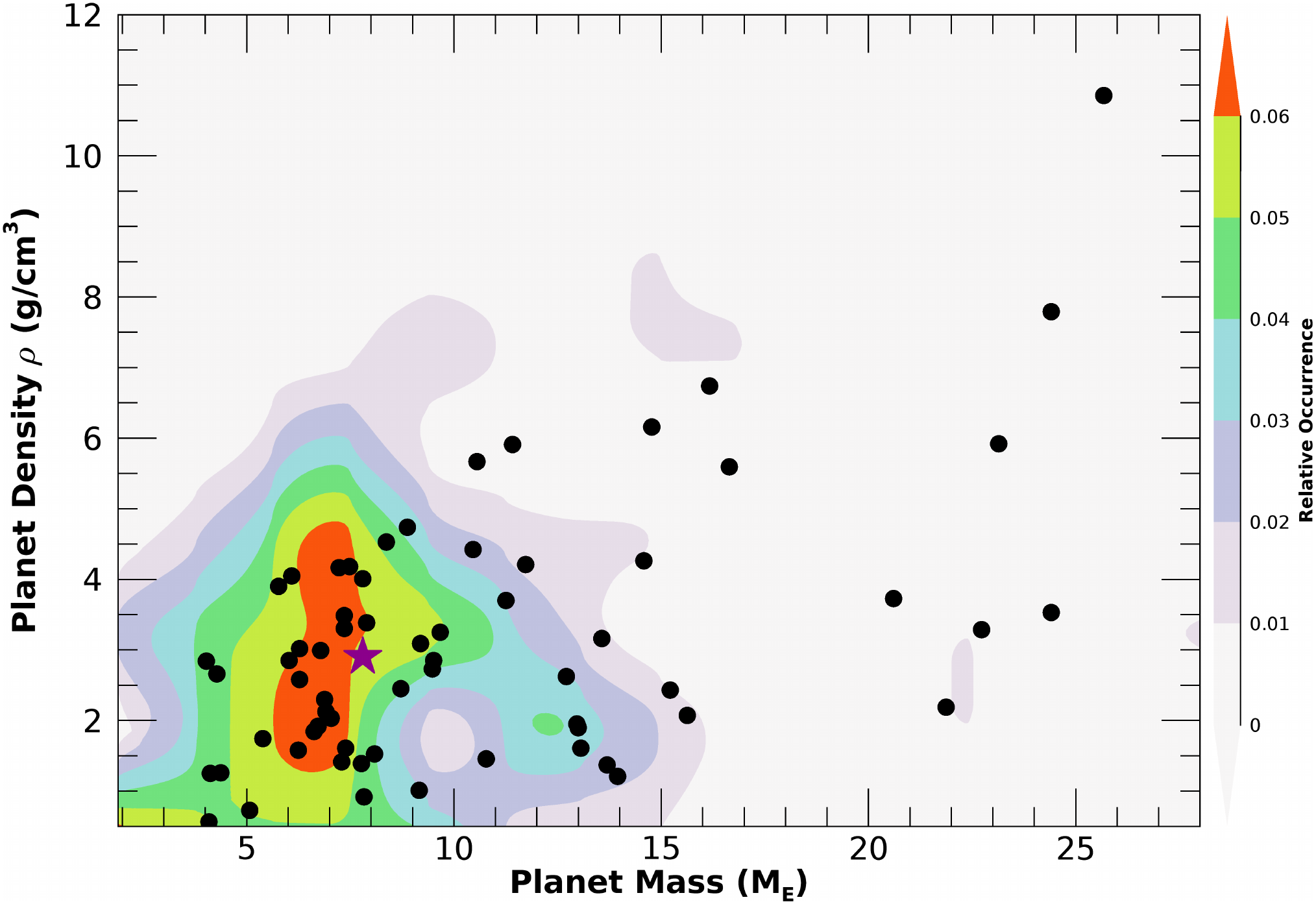} \\
\end{tabular}
    \caption{Top left: mass-radius diagram with contours built using the exoplanets sample from Fig. \ref{mrrelation}. Black dots represent planets with mass measurements at the $>3\sigma$ level of statistical significance. HD 5278 b is indicated by a large dark magenta star. The contours are color-coded in terms of relative occurrence. Top right: The same plot in the planetary mass - stellar irradiation plane. Bottom left: the same plot in the planetary density vs. stellar irradiation plane. Bottom right: the same plot in the planetary density vs. mass plane.}
    \label{contourplots}
\end{figure*}

\begin{figure}
    \centering
    \includegraphics[width=0.49\textwidth, angle=0]{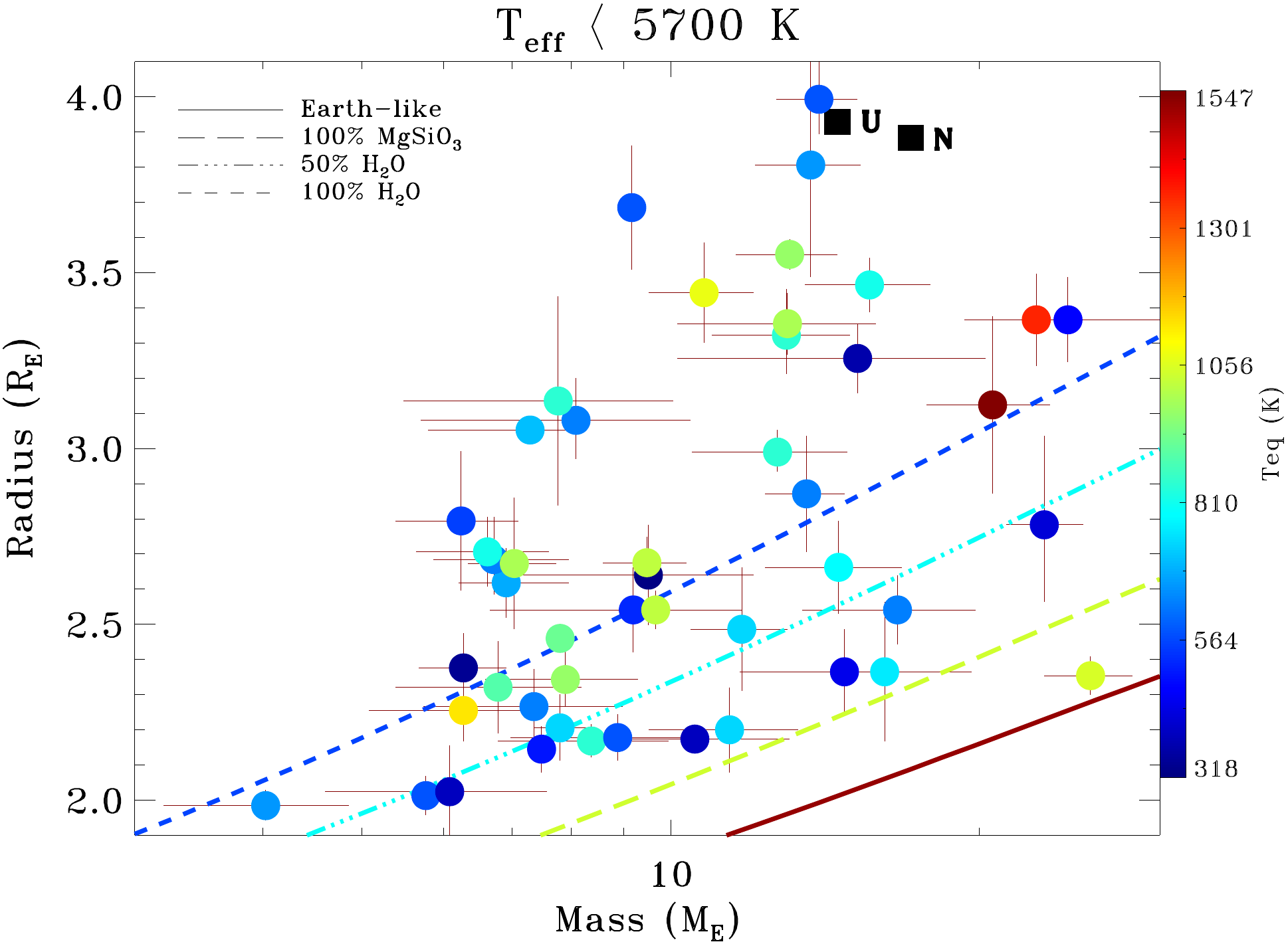} \\
    \includegraphics[width=0.49\textwidth, angle=0]{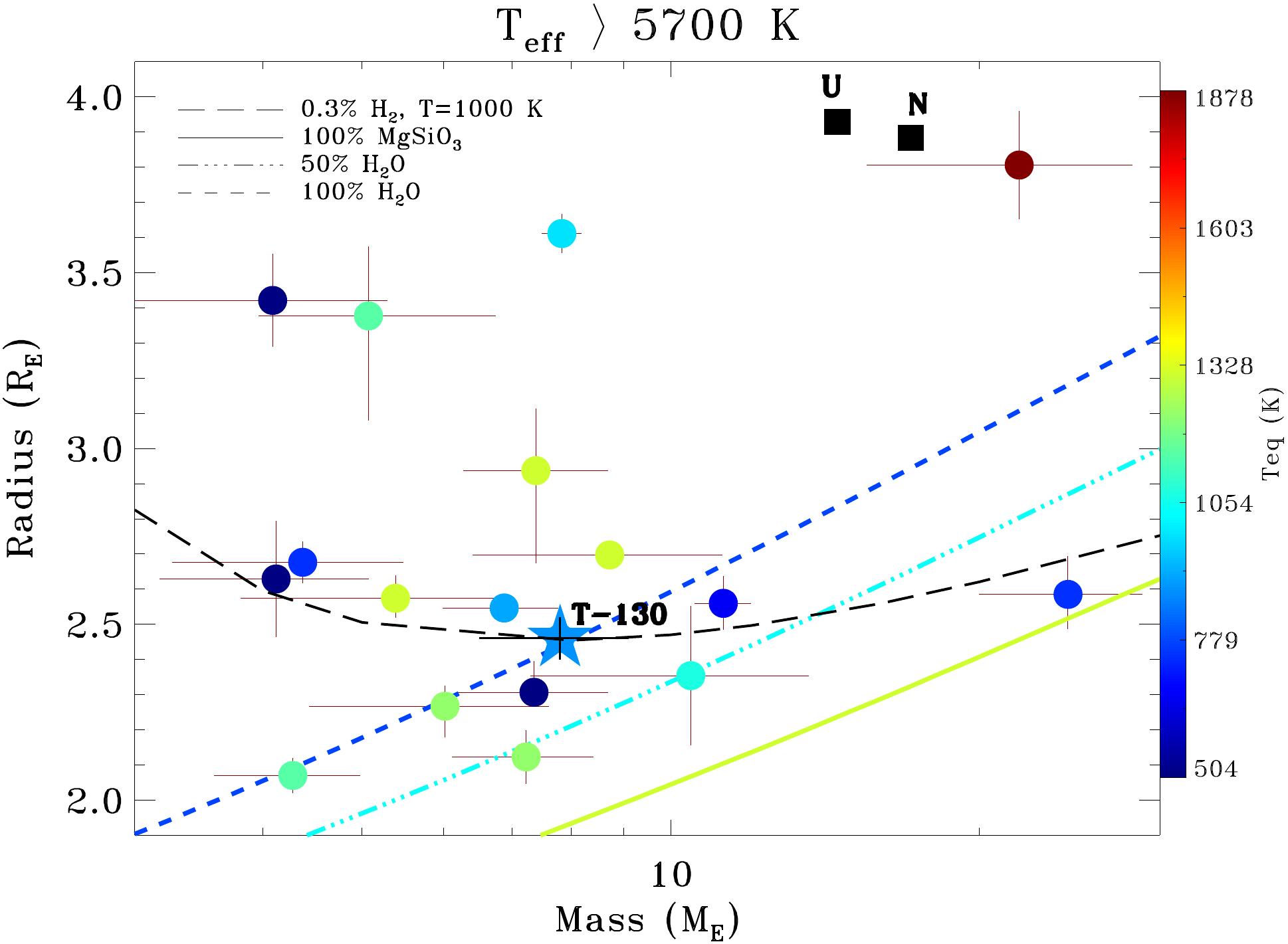} \\
\caption{Top: Mass-radius diagram of sub-Neptunes with masses determined at the $3\sigma$ level (or better), and with T$_\mathrm{eff}<5700$ K. Color-code and curves are from Figure \ref{mrrelation}. Bottom: The same plot for the sample with T$_\mathrm{eff}>5700$ K, including HD 5278 b (star).}
    \label{mrdiag_var_teff}
\end{figure}

\begin{figure}
    \centering
    \includegraphics[width=0.49\textwidth, angle=0]{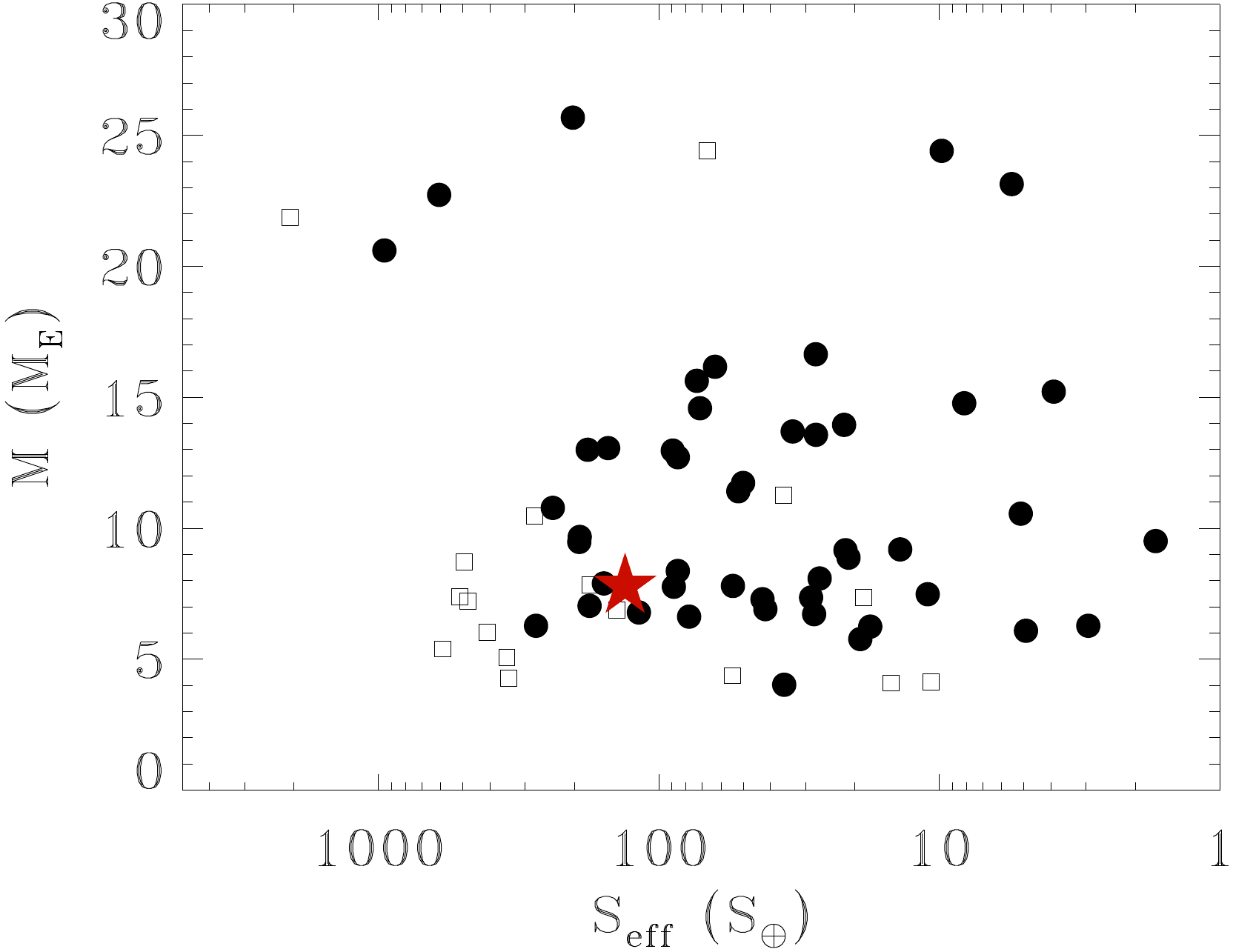} \\
    \includegraphics[width=0.49\textwidth, angle=0]{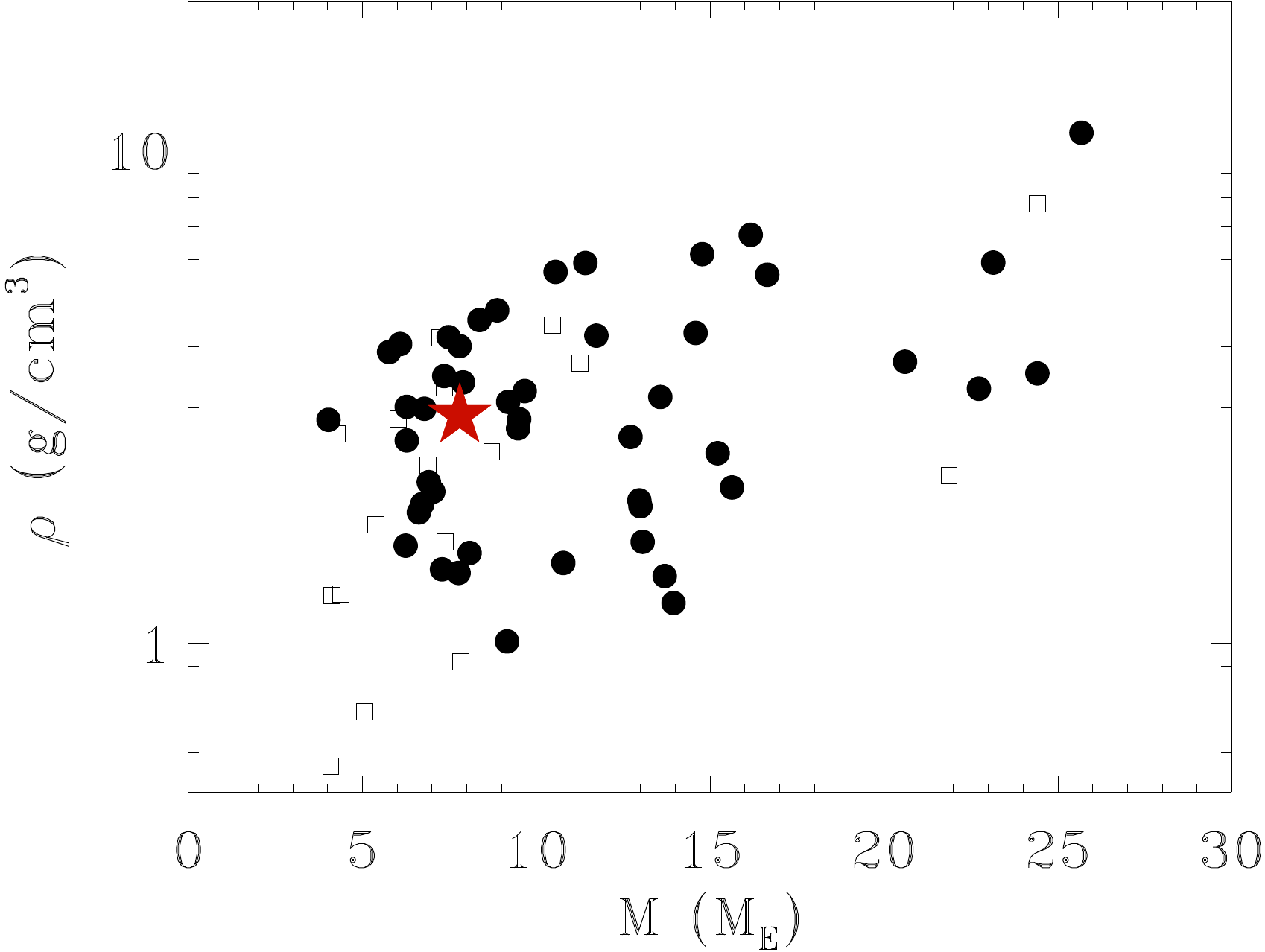} \\
    \caption{Top: planet mass vs. stellar irradiation for the sample defined in Figure \ref{mrdiag_var_teff}, including HD 5278 b (red star). Filled dots and open squares indicate objects around primaries with T$_\mathrm{eff}<5700$ K and T$_\mathrm{eff}>5700$ K, respectively. Bottom: same as the top panel, this time in the planet density vs. planet mass diagram.}
    \label{mass_rho_seff}
\end{figure}

\begin{table}[h!]
\centering
\caption{Results of the K-S on different subsets of the mass distribution of transiting sub-Neptunes: cut-offs at T$_\mathrm{eff}=5400$ K (1), T$_\mathrm{eff}=5500$ K (2), T$_\mathrm{eff}=5600$ K (3), T$_\mathrm{eff}=5700$ K (4), T$_\mathrm{eff}=5800$ K (5), and T$_\mathrm{eff}=5900$ K (6).}
\label{kstest}
\begin{tabular}{@{}cccc}
\hline
Sub-samples & $Pr(D)$ & $N_{\mathrm{low}}$ & $N_{\mathrm{high}}$\\
\hline
1 & 0.38 & 37 & 25 \\
2 & 0.11 & 40 & 22 \\
3 & 0.02 & 43 & 19 \\
4 & 0.02 & 47 & 15 \\
5 & 0.02 & 50 & 12 \\
6 & 0.04 & 52 & 10 \\
\hline
\end{tabular}
\end{table}

As mentioned in the Introduction, high-precision mass determinations, at the 20\% level or better, have recently been recognized as mandatory in order to carry out detailed atmospheric characterization studies \citep{Batalha2019}. Inferences on both mass and chemical composition of the atmosphere based on transmission/emission spectra alone \citep{deWit2013} might be dominated by strong degeneracies, particularly in the case of sub-Neptunes and rocky super Earths \citep{Batalha2017}. It is therefore critical to augment the number of sub-Neptunes with well-determined dynamical mass estimates (Figure \ref{mrrelation}). 

The growing sample of such objects can be investigated in a statistical sense in order to uncover patterns (e.g., overdensities) in parameter space that might represent fossil evidence of the planets' formation and evolution history, as it has been done recently using radius alone \citep{Fulton2017,Fulton2018}. Considering all sub-Neptunes with masses detected with $\geq 3\sigma$ confidence, as we did in Figure \ref{mrrelation}, the sample size (62 objects) is indeed not yet very large, so the findings we present in the following are to be considered illustrative. Given the limited statistics available, the preparation of the contour/density plots shown in the remainder of this Section is based on a simple binned histograms analysis rather than the adoption of more sophisticated approaches based on e.g. (weighted) kernel density estimators \citep{Morton2014, Fulton2017}. 

The upper left panel of Figure \ref{contourplots} shows how close-in sub-Neptunes with $5\lesssim M_p\lesssim 10$ M$_\oplus$ and $2\lesssim R_p\lesssim 3.0$ R$_\oplus$ appear intrinsically more common than planets with higher masses and radii. The possibly double peaked radius distribution in this mass regime (at $\sim2.1-2.3$ R$_\oplus$ and $\sim2.5-2.7$ R$_\oplus$) might well be an artifact due to the binned histogram approach utilized to derive the density plots, in the face of small-number statistics. Interestingly, however, a hint of the presence of this feature is also seen in Figures 6 and 8 of \citet{Fulton2018}. We also note that planets in the larger radius range $R_p\gtrsim 3.0$ R$_\oplus$ appear uniformly distributed in the full mass range, and similarly planets in the higher mass range $10\lesssim M_p\lesssim 15$ M$_\oplus$ seem to be uniformly distributed over the whole $2<R_p<4$ R$_\oplus$ range. 

The upper right panel of Figure \ref{contourplots} shows instead that sub-Neptunes with $5<M_p<10$ M$_\oplus$ are found over a large interval of irradiation levels ($10\lesssim S_\oplus\lesssim 250$), with a narrow peak in relative occurrence of objects with $M_p\sim 7$ M$_\oplus$ in the $30\lesssim S_\oplus\lesssim 100$ range. The more massive planets with $10<M_p<15$ M$_\oplus$ appear to be slightly more abundant in conditions of higher irradiation (S$_\oplus\gtrsim 30$), with a clear majority of the highest-mass sub-Neptunes ($M_p> 20$ M$_\oplus$) receiving $S_\oplus\gtrsim 100$. 

In the bottom left panel of Figure \ref{contourplots} we notice how sub-Neptunes with densities in the range $\varrho\sim 2.0-3.0$ g cm$^{-3}$  are distributed over almsot three orders of magnitude in stellar irradiation levels ($3\lesssim S_\oplus\lesssim 2000$), with a peak in relative occurrence in a somewhat narrow regime of insolation ($50\lesssim S_\oplus \lesssim 100$). The lowest-density objects appear instead distributed mostly over the $S_\oplus \sim10-100$ range. The highest-density objects ($\varrho\gtrsim 6.0$ g cm$^{-3}$) mostly reside at $S_\oplus\gtrsim 50$.

Finally, in the bottom right panel of Figure \ref{contourplots} it appears that sub-Neptunes with $5<M_p<10$ M$_\oplus$ can span an important range of densities ($1\lesssim\varrho\lesssim 5$ g cm$^{-3}$), the peak in relative occurrence encompassing objects with $6\lesssim M_p\lesssim8$ M$_\oplus$ and $2\lesssim \varrho\lesssim 4$. Higher-mass ($10<M_p<15$ M$_\oplus$) sub-Neptunes with $\varrho\lesssim 3$ g cm$^{-3}$ appear intrinsically slightly more common, although the highest-mass sub-Neptunes also populate the tail of densest systems. 

While one should exert caution not to over-interpret the patterns of variations in relative occurrence in parameter space presented in Figure \ref{contourplots}, given the still relatively small-number statistics we are dealing with, it is certainly possible to provide a few (more or less speculative) comments in connection to planet formation and evolution scenarios. For example, the overdensity in the upper left panel of Figure \ref{contourplots} is not unexpected, given the known behaviour of planet frequency as a function of radius and mass for small planets \citep{Mayor2011,Howard2012}. It is tempting to try to explain the observed double peak in the radius and mass ranges $\sim2.1-2.3$ R$_\oplus$ - $\sim5.0-8.0$ M$_\oplus$ and $\sim2.5-2.7$ R$_\oplus$, $\sim7.0-10.0$ M$_\oplus$, respectively, as evidence of intrinsic compositional differences: the former peak could correspond to the high-mass tail of a population of truly rocky planets (or evaporated cores), while the latter peak could be made up of objects with icy cores and/or small gaseous envelopes (e.g., \citealt{Venturini2020,Neil2020}). 

Interestingly, objects with $\sim7$ M$_\oplus$ appear intrinsically more common over an important range of irradiation levels and densities (top right and bottom right panels of Figure \ref{contourplots}, respectively). The shape of the distribution of measured masses for this sample of sub-Neptunes closely resembles that realized by planet formation and evolution simulations in the recent work of \citet{Venturini2020} that produce icy cores ($\sim50\%$ in mass), although the peak in the distribution in their analysis is shifted to slightly higher masses (9-10 M$_\oplus$). This might further corroborate the notion that an important fraction of sub-Neptunes populating the second peak in the radius valley might be composed of water-rich planets with thin H$_2$/He atmospheres, rather than truly rocky planets surrounded by H$_2$/He envelopes with mass fractions of a few percent, as predicted by pure photoevaporation models \cite{Owen2017} and core-powered mass-loss models \cite{Ginzburg2018}. 

The mild trend of more massive / higher density sub-Neptunes being more common in conditions of higher incident fluxes seen in the upper right and bottom left panels of Figure \ref{contourplots} can be qualitatively understood within the known paradigms of planet formation and migration when effects due to evaporation are included. In particular, for low-mass sub-Neptunes, the dominant effect of intense irradiation is atmospheric escape, which increases the planetary density such that $\varrho$ and $S_\oplus$ are anti-correlated (e.g., \citealt{Jin2018}).

High-mass (M$_p>15$ M$_\oplus$), high-density sub-Neptunes are rare (see bottom left and right panels of Figure \ref{contourplots}). They typically receive irradiation levels significantly lower than those characterizing the evaporation desert (as defined by \citealt{Lundkvist2016}), therefore they are not likely to have been stripped of their envelopes. This sparse population likely carries the imprint of evolutionary pathways that might include original high-density formation and subsequent inward migration via mechanisms involving multiple companions (e.g., Kozai-Lidov oscillations. See for example \citealt{Dawson2014,Mustill2017}), processes inhibiting gas accretion or promoting gas removal after formation (e.g., \citealt{Ormel2015,Guilera2020}), or a possible history of impacts in the early stages of the formation process \citep{Ogihara2020,Venturini2020}, for which only recently direct observational evidence has been produced \citep{Santerne2018,Bonomo2019}. 

The mass-radius diagram of sub-Neptunes shows another intriguing feature. The two panels of Figure \ref{mrdiag_var_teff} highlight a dearth of $M_p > 10$ M$_\oplus$ companions around G- and F-type stars with T$_\mathrm{eff}>5700$ K, particularly for objects with $R_p\gtrsim2.5$ R$_\oplus$\footnote{Interestingly, the only companion with $M_p>10$ M$_\oplus$ and a radius comparable to that of Uranus and Neptune is K2-100b, orbiting on a very short period a young late-F-type star. Its radius is expected to shrink even below 3 R$_\oplus$ over the main-sequence lifetime of the star \citep{Barragan2019}}. A Kolmogorov-Smirnov (K-S) test returned a 2.4\% probability that the mass distributions of sub-Neptunes around primaries with T$_\mathrm{eff}<5700$ K and T$_\mathrm{eff}>5700$ K are the same. This feature is seen across a wide range of insolation levels (see top panel of Figure \ref{mass_rho_seff}). The two populations are also quite clearly identified in a $\varrho$-M$_p$ diagram (bottom panel of Figure \ref{mass_rho_seff}). The feature does not seem to be produced by an observational bias, as at a given value of orbital period it is the less massive companion the one that is harder to detect around an increasingly more massive primary. The difference in mass distribution vanishes if a cutoff at slightly cooler temperatures is applied (T$_\mathrm{eff}>5400$ K). A summary of K-S test results is provided in Table \ref{kstest}, in which we do not investigate the differences in the subsets of the distributions with cutoffs at T$_\mathrm{eff}>5900$ K as the sample size of the distribution for hotter stellar hosts drops below the number ($\sim10$) for which the K-S test provides reliable results. 

The hint of a lack of close-in, higher-mass (M$\gtrsim 10$ M$_\oplus$) sub-Neptunes around G-F primaries can be interpreted as fossil evidence of planet formation around stars of varied mass. More massive primaries typically have higher-mass disks. Upon reaching the critical mass (10 M$_\oplus$ or so), newly formed cores have a higher chance of quickly accreting large amount of gas, ending up their formation process as giant planets. The threshold appears to be for planets around primaries with T$_\mathrm{eff}\gtrsim 5500$ K. 

\section{Summary and Conclusions}

We performed high-precision RV follow-up with ESPRESSO for mass confirmation of the transiting sub-Neptune candidate TOI-130 b, uncovered by TESS orbiting the bright, nearby, late-F dwarf HD 5278 with a period $P_b=14.34$ days. Our main findings can be summarized as follows: 

$a)$ The combination of TESS photometry and ESPRESSO RVs allowed us to determine a planetary mass and radius $M_{\rm b} = 7.8_{-1.4}^{+1.5}$ M$_\oplus$ and $R_{\rm b} = 2.45\pm0.05$ R$_\oplus$, respectively. The resulting planetary bulk density is $\varrho_{\rm b} = 2.9_{-0.5}^{+0.6}$ g cm$^{-3}$. Our ESPRESSO RVs unveiled the presence of a second, non-transiting companion in the system, for which we measured a period $P_c=40.87_{-0.17}^{+0.18}$ days and determined a minimum mass $M_c\sin i_c =18.4_{-1.9}^{+1.8}$ M$_\oplus$. 

$b)$ HD 5278 b can be described as a water-rich ($\sim30\%$ in mass) sub-Neptune, with a gaseous envelope that comprises $\sim0.4\%$ and $\sim17\%$ of its total mass and radius, respectively. A 'dry' model with small rocky core, a large silicate mantle, and a sizable H$_2$/He envelope is also compatible with the measured physical properties of the planet. Given the system parameters, HD 5278 b represents one of the best targets orbiting G-F primaries for follow-up atmospheric characterization measurements with HST and JWST, which could be instrumental in resolving compositional degeneracies. 

$c)$ The HD 5278 b,c planetary system, with a period ratio not particularly close to a resonant configuration and low eccentricities, is Hill-stable even in the case of highly non-coplanar orbits. On the one hand, such a scenario is possible in light of the very low levels of stellar activity recorded, lack of detectable rotation in both RVs and photometry and low $v\sin i$ value, conditions that, given the spectral type of the primary, are statistically more likely if the star is viewed not far from pole-on. On the other hand, the lack of detectable eccentricity for both planets hints at the fact that their orbits might not be particularly misaligned. Measurements of the spin-orbit (mis)alignment for HD 5278 b would be telling. 

$d)$ We placed the properties of HD 5278 b in the context of the (still small) presently-known sample of sub-Neptunes with measured mass and radius. We showed that the lower-mass, smaller-radius component of the population is intrinsically more common, with a possibly double-peak radius distribution bracketing 2.5 R$_\oplus$ and a mass distribution clearly peaking at $\sim7$ M$_\oplus$. We highlighted a possible excess of higher-mass, higher-density  sub-Neptunes in conditions of stronger irradiation. We provided evidence that very high-density, higher-mass sub-Neptunes are intrinsically rare. These features can be understood in broad terms within the context of modeling efforts of planetary systems formation processes and the subsequent orbital evolution.

$e)$ We identified a marginally significant lack of more massive sub-Neptunes around G-F primaries with T$_\mathrm{eff}\gtrsim 5500$ K, when compared to the mass distribution of these objects around cooler primaries (probability that the two distributions are the same of 2\%). This is the likely imprint of a disk mass dependence in the outcome of sub-Neptune planet formation that we capture today in terms of a different mass distribution of close-in sub-Neptunes around stars of varied spectral type. 

In conclusion, the results of our combined ESPRESSO+TESS analysis of the HD 5278 planetary system reinforce the importance of the detailed, high-precision determination of the fundamental physical parameters of transiting sub-Neptunes, particularly those orbiting solar-type primaries. With increasing sample sizes of the population with well-determined properties, the preliminary evidence for trends in parameter space (planet mass and radius, bulk density, stellar insolation and effective temperature) reported here will be put on firmer statistical grounds. It will then become possible to compare not only qualitatively but also quantitatively theoretical predictions of planet formation and evolution models with the observed population of sub-Neptunes, performing at higher resolution key diagnostic studies of e.g. the detailed shape of the combined mass and radius distribution, or trends of occurrence rates and bulk composition with spectral type of the primary and irradiation levels/orbital separation. 

\begin{acknowledgements}

The authors acknowledge the ESPRESSO project team for its effort and dedication in building the ESPRESSO instrument. This work has received financial support from the ASI-INAF agreement n.2018-16-HH.0. M.D. acknowledges financial support from the FP7-SPACE Project ETAEARTH (GA No. 313014).
The INAF authors acknowledge financial support of the Italian Ministry of Education, University, and Research
with PRIN 201278X4FL and the "Progetti Premiali" funding scheme.
The ESPRESSO Instrument Project was partially funded through
SNSF’s FLARE Programme for large infrastructures. This work has been
carried out in part within the framework of the NCCR PlanetS supported by
the Swiss National Science Foundation. This work was supported by FCT
- Funda\,c\~ao para a Ci\^encia e Tecnologia through national funds and by
FEDER through COMPETE2020 - Programa Operacional Competitividade e
Internacionaliza\,c\~ao by these grants: UID/FIS/04434/2019; UIDB/04434/2020;
UIDP/04434/2020; PTDC/FIS-AST/32113/2017 \ POCI-01-0145-FEDER-
032113; PTDC/FIS-AST/28953/2017 \& POCI-01-0145-FEDER-028953;
PTDC/FIS-AST/28987/2017 \& POCI-01-0145-FEDER-028987; PTDC/FIS-OUT/29048/2017 \& IF/00852/2015. 
S.C.C.B. acknowledges support from FCT through contract nr. IF/01312/2014/CP1215/CT0004.
S.G.S acknowledges the support from FCT through Investigador FCT contract nr. CEECIND/00826/2018 and POPH/FSE (EC).
This project has received funding from the European Research Council (ERC) under the
European Union’s Horizon 2020 research and innovation programme (project Four Aces, grant agreement No 724427). V.A. acknowledges the support from FCT through Investigador FCT contract nr. IF/00650/2015/CP1273/CT0001. 
Y.A. and J.H. acknowledge the Swiss National Science Foundation (SNSF) for supporting research through the SNSF grant 200020\_192038. 
J.I.G.H. acknowledges financial support from Spanish Ministry of
Science and Innovation (MICINN) under the 2013 Ram\'on y Cajal 
programme RYC-2013-14875. 
J.I.G.H., A.S.M., R.R., and C.A.P. acknowledge financial support 
from the Spanish MICINN AYA2017-86389-P.
A.S.M. acknowledges financial support from the Spanish Ministry of
Science and Innovation (MICINN) under the 2019 Juan de la Cierva
Programme. 
R. A. is a Trottier Postdoctoral Fellow and acknowledges support from the Trottier Family Foundation.This work was supported in part through a grant from FRQNT.
This work has made use of data from the European Space Agency (ESA) mission {\it Gaia} (\url{https://www.cosmos.esa.int/gaia}), processed by the {\it Gaia} Data Processing and Analysis Consortium (DPAC,
\url{https://www.cosmos.esa.int/web/gaia/dpac/consortium}). Funding
for the DPAC has been provided by national institutions, in particular
the institutions participating in the {\it Gaia} Multilateral Agreement. This publication makes use of The Data \& Analysis Center for Exoplanets (DACE), which is a facility based at the University of Geneva (CH) dedicated to extrasolar planets data visualisation, exchange and analysis. DACE is a platform of the Swiss National Centre of Competence in Research (NCCR) PlanetS, federating the expertise in Exoplanet research. The DACE platform is available
at \url{https://dace.unige.ch}. This paper includes data collected by the TESS mission, which are publicly available from the Mikulski Archive for Space Telescopes (MAST). We acknowledge the use of public TESS Alert data from pipelines at the TESS Science Office and at the TESS Science Processing Operations Center. Resources supporting this work were provided by the NASA High-End Computing (HEC) Program through the NASA Advanced Supercomputing (NAS) Division at Ames Research Center for the production of the SPOC data products.

\end{acknowledgements}

%
%

   \bibliographystyle{aa} 
   \bibliography{TOI-130_biblio} 



\begin{appendix} 

\section{1D Coadded ESPRESSO Spectrum of HD 5278}

\begin{figure*}
\includegraphics[width=0.98\textwidth]{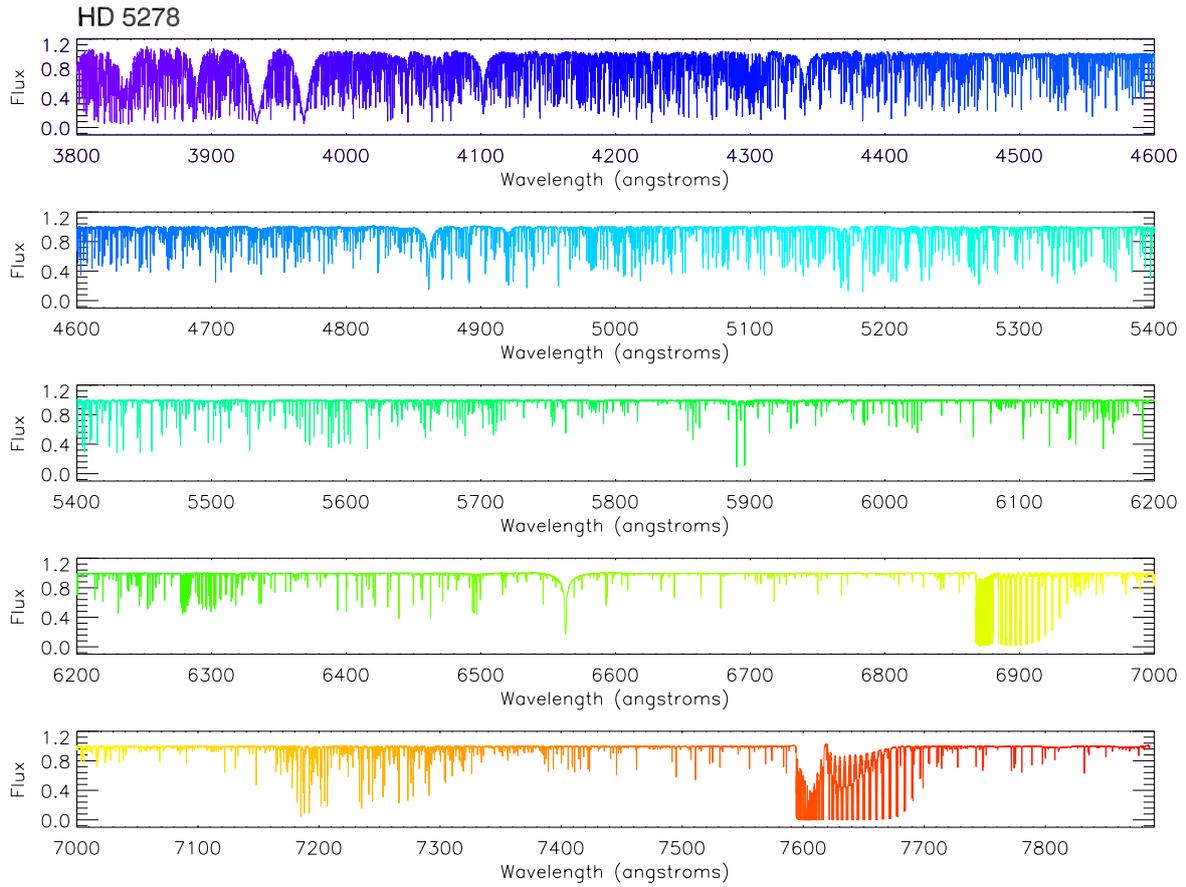}
\caption{Co-added, normalized, merged, and RV-corrected ESPRESSO spectrum of HD 5278 produced using the StarII DAS workflow on the available set of 43 2D ESPRESSO spectra of the target.}
\label{1d_spectrum}
\end{figure*}

\newpage

\section{Radial velocities, activity indicators,  and posterior distributions}
\begin{figure*}
\includegraphics[width=0.98\textwidth]{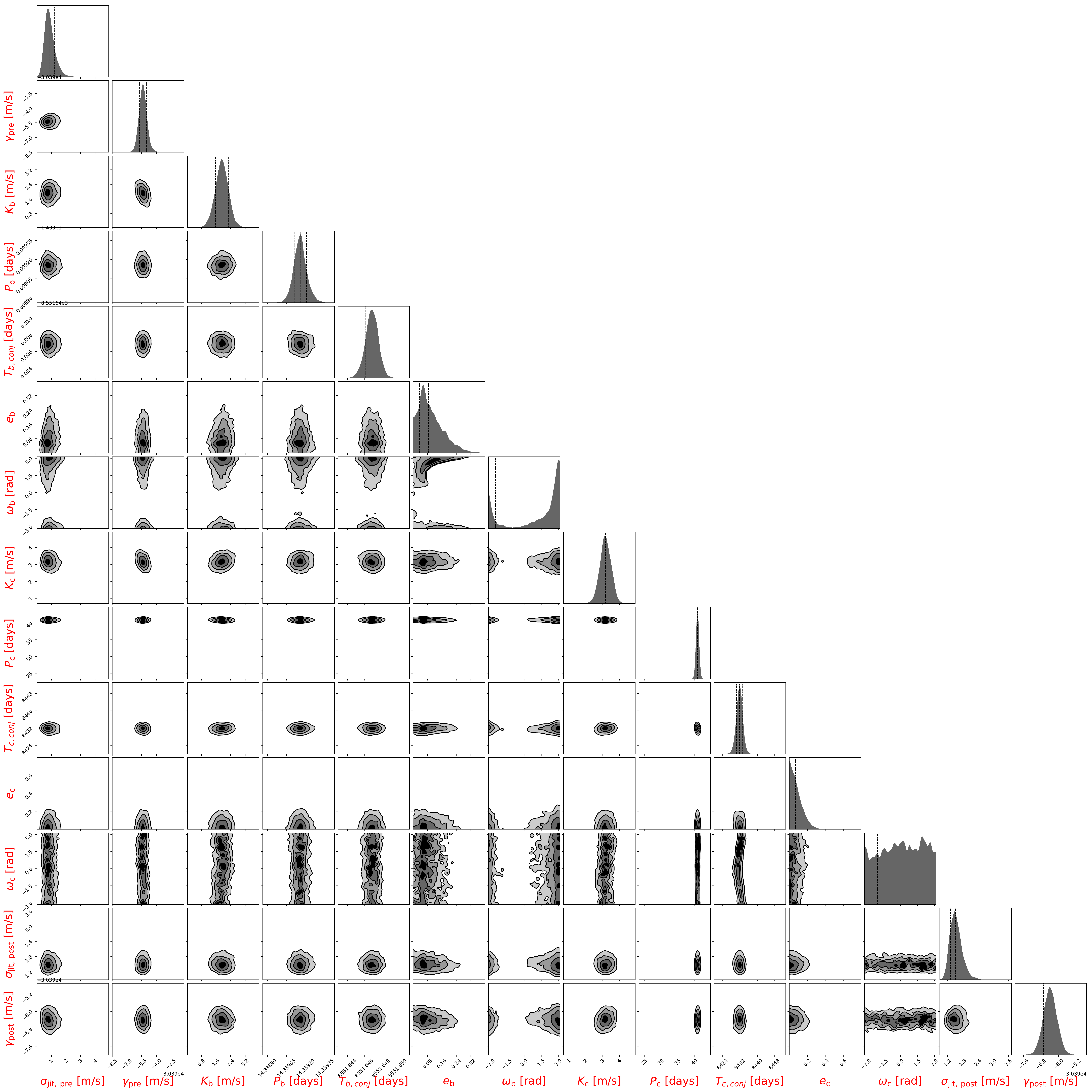}
\caption{Posterior distributions of the photometry-related system parameters in the joint ESPRESSO RV + TESS photometry analysis.}
\label{fig:posteriors_phot}
\end{figure*}

\begin{figure*}
\includegraphics[width=0.98\textwidth]{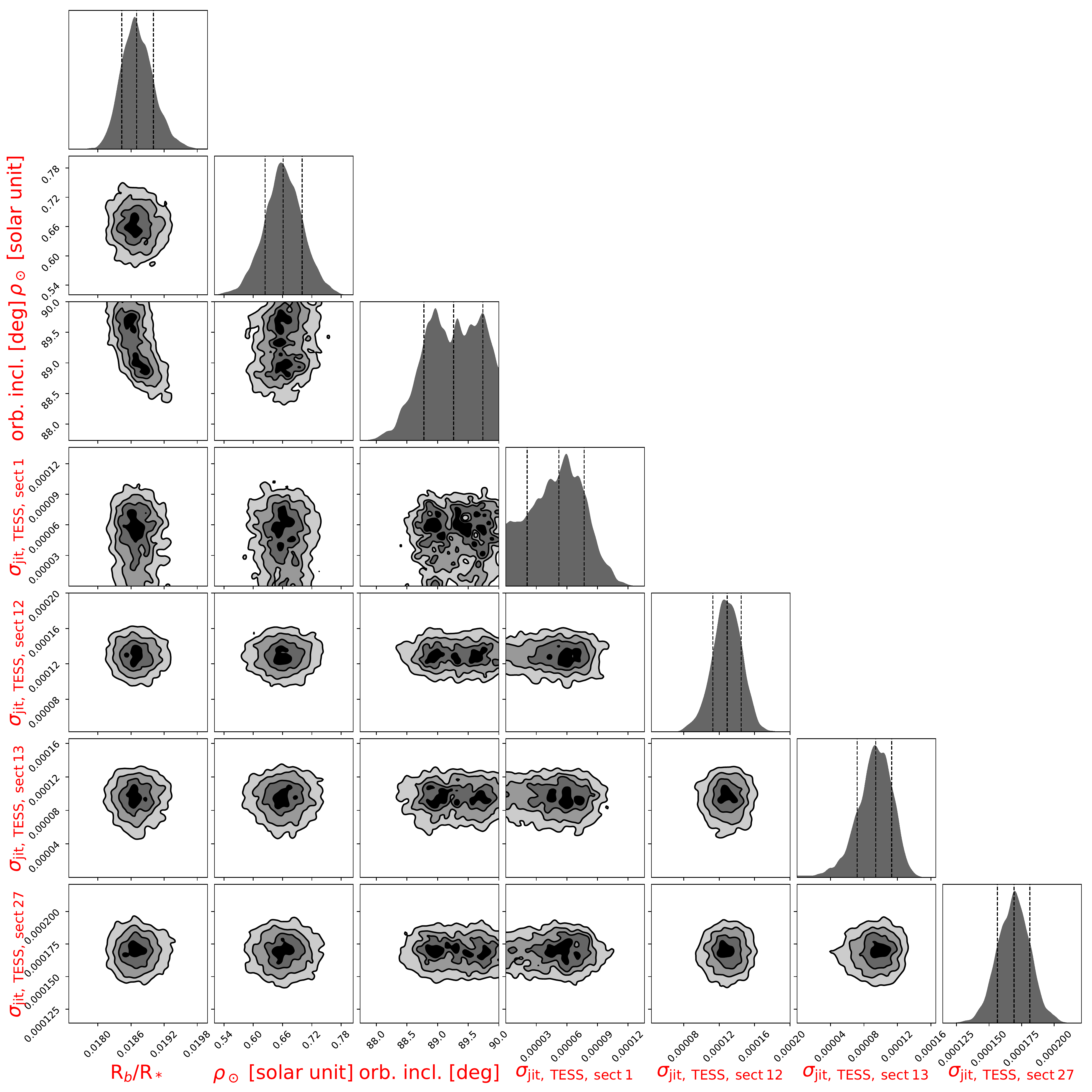}
\caption{Posterior distributions of the spectroscopy-related system parameters in the joint ESPRESSO RV + TESS photometry analysis.}
\label{fig:posteriors_spec}
\end{figure*}

\begin{figure*}
\includegraphics[width=0.98\textwidth]{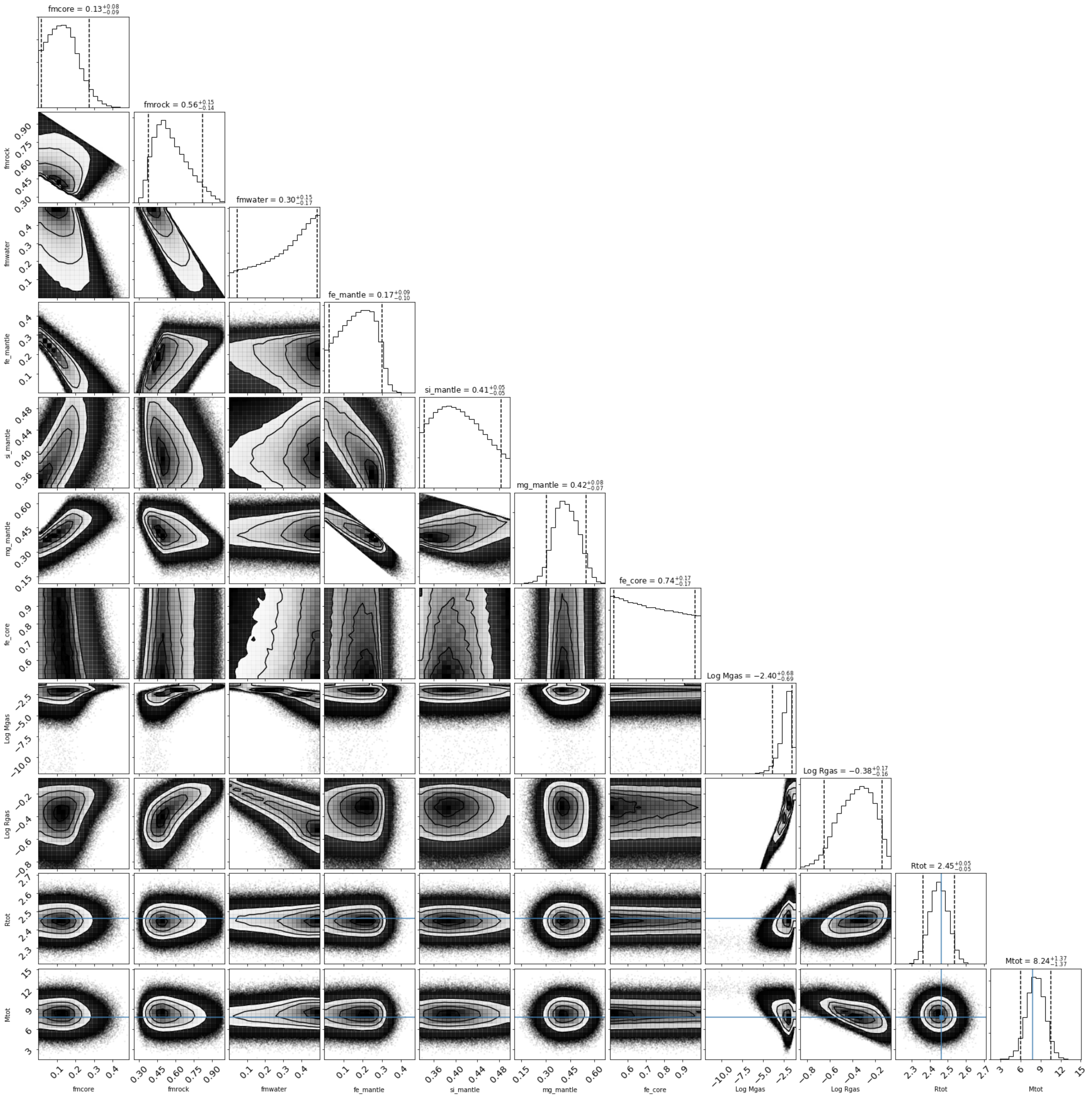}
\caption{Corner plot showing the results of the internal structure retrieval for HD 5278 b in the case of a fully differentiated planet model. The different columns refer to, respectively, the core (iron + sulfur) mass fraction, the silicate mantle mass fraction, the water mass fraction, the molar fraction of Fe, Si and Mg in the mantle, the molar fraction of Fe in the core, the $\log$ of the gas mass (in Earth masses), the $\log$ of the atmospheric thickness (Earth radii), the total radius and mass. In the two last columns, the blue lines show the observed values. For each column, the mean of the distribution as well as the 5\% and 95\% quantile are indicated on the top, and the 5\% and 95 \% quantiles are also indicated with dashed vertical lines.}
\label{fig:posteriors_model1}
\end{figure*}

\begin{figure*}
\includegraphics[width=0.98\textwidth]{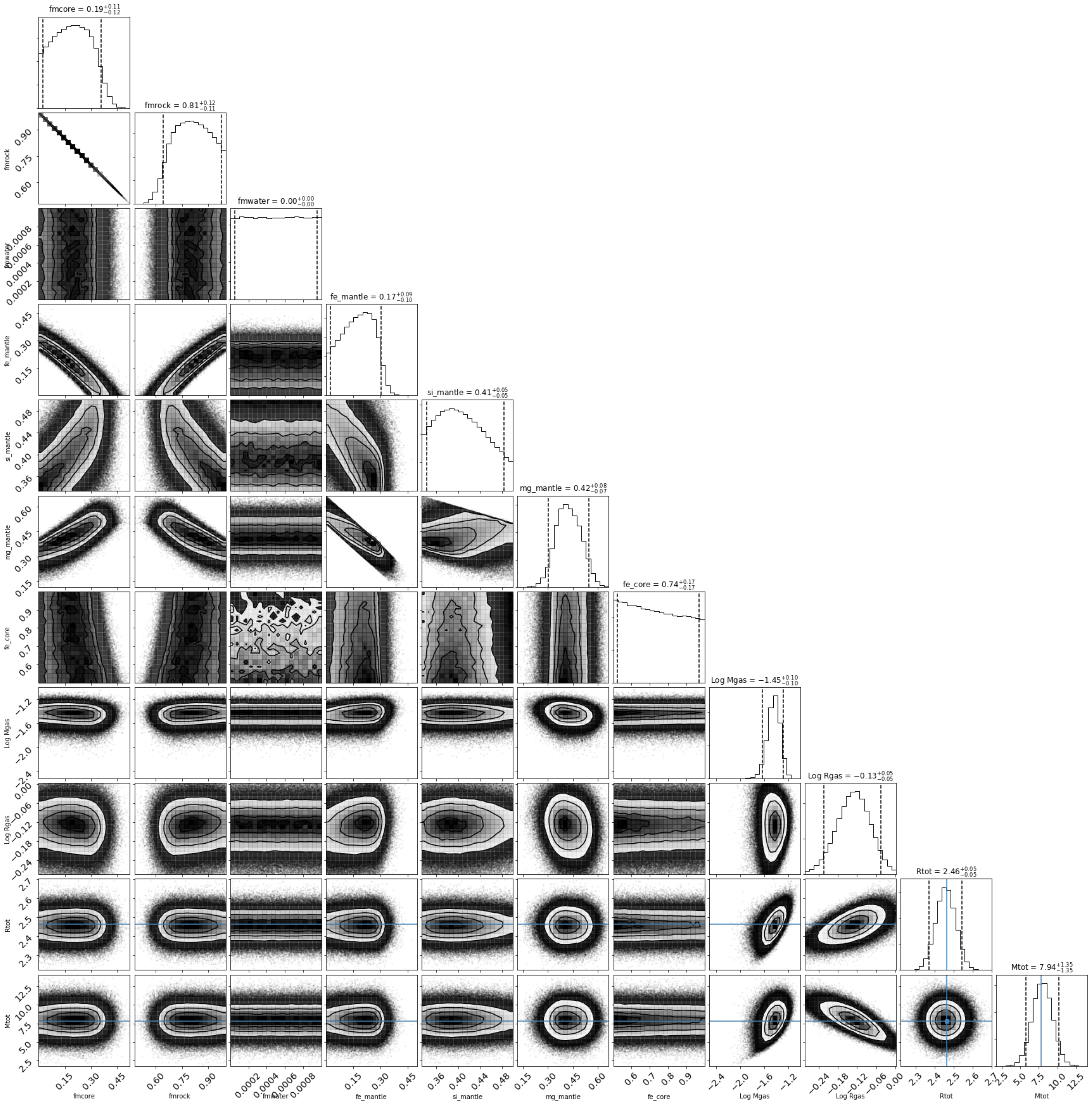}
\caption{same as Figure \ref{fig:posteriors_model1} for the 'dry' model.}
\label{fig:posteriors_model2}
\end{figure*}

Table \ref{tab:rv} contains the ESPRESSO RV data and their uncertainties, Table \ref{tab:activ} lists the time-series of spectroscopic activity indicators. Figures \ref{fig:posteriors_phot} and \ref{fig:posteriors_spec} show the posterior distributions of all the model parameters in the joint fit to the ESPRESSO RV data and TESS light curve. Figures \ref{fig:posteriors_model1} and \ref{fig:posteriors_model2} show corner plots of the posterior distributions of the relevant parameters of the two interior structure models. 

\begin{table*}
\centering
\caption{ESPRESSO radial velocities of HD 5278. In the last column, PRE and POST indicate spectra gathered before and after the intervention (see text).}            
\setlength{\tabcolsep}{1.0mm}
\renewcommand{\footnoterule}{}                          
\begin{tabular}{c c c c}        
\hline \hline
\noalign{\smallskip}
$ \rm BJD_{UTC}$ & RV & $\pm 1~\sigma$ & Case \\
$-2400000$   & ($\ms$) & ($\ms$) &  \\
\noalign{\smallskip}
\hline
\noalign{\smallskip}
 58416.564987 &  $-30390.87$ &       0.61 &   PRE \\
 58420.623837 &  $-30390.01$ &       0.98 &   PRE \\
 58421.615745 &  $-30390.98$ &       0.37 &   PRE \\
 58426.662776 &  $-30394.76$ &       0.36&   PRE \\
 58432.512969 &  $-30393.95$ &       0.42&   PRE \\
 58459.511927 &  $-30393.04$ &       0.39&   PRE \\
 58466.533613 &  $-30393.95$ &       0.47&   PRE \\
 58477.521619 &  $-30396.33$ &       0.58&   PRE \\
 58488.535606 &  $-30396.38$ &       0.36&   PRE \\
 58490.544982 &  $-30395.07$ &       0.51&   PRE \\
 58686.921063 &  $-30398.91$ &       0.37&   POST \\
 58688.776700 &  $-30396.79$ &       0.66&  POST \\ 
 58688.788018 &  $-30396.99$ &       0.66&  POST \\
 58706.867055 &  $-30389.82$ &       0.32&  POST \\
 58719.844569 &  $-30392.52$ &       0.28&  POST \\
 58721.863285 &  $-30398.29$ &       0.30&  POST \\
 58725.809120 &  $-30400.87$ &       0.53&  POST \\
 58731.758304 &  $-30397.86$ &       0.45&  POST \\
 58737.814370 &  $-30395.17$ &       0.38&  POST \\
 58741.624983 &  $-30395.79$ &       0.33&  POST \\
 58753.617398 &  $-30395.60$ &       0.28&  POST \\
 58759.665785 &  $-30397.48$ &       0.44&  POST \\
 58761.674737 &  $-30395.95$ &       0.44&  POST \\
 58766.630575 &  $-30401.10$ &       0.36&  POST \\
 58768.626690 &  $-30402.71$ &       0.33&  POST \\
 58771.609041 &  $-30402.54$ &       0.46&  POST \\
 58773.688306 &  $-30397.31$ &       0.38&  POST \\
 58776.572183 &  $-30398.68$ &       0.38&  POST \\
 58776.694342 &  $-30397.16$ &       0.51&  POST \\
 58778.631061 &  $-30394.20$ &       0.32&  POST \\
 58780.645934 &  $-30394.84$ &       0.26&  POST \\
 58782.560951 &  $-30397.76$ &       0.39&  POST \\
 58784.531509 &  $-30397.10$ &       0.57&  POST \\
 58786.600937 &  $-30393.26$ &       0.30&  POST \\
 58788.597405 &  $-30394.11$ &       0.34&  POST \\
 58792.609247 &  $-30393.27$ &       0.38&  POST \\
 58801.619078 &  $-30397.47$ &       0.36&  POST \\
 58811.617055 &  $-30401.00$ &       0.49&  POST \\
 58815.611406 &  $-30396.80$ &       0.33&  POST \\
 58817.547039 &  $-30396.56$ &       0.33&  POST \\
 58819.554583 &  $-30396.31$ &       0.65&  POST \\
\noalign{\smallskip}
\hline \hline
\label{tab:rv}
\end{tabular}
\end{table*}

\begin{table*}
\centering
\caption{Spectroscopic activity indicators for HD 5278 (see main text).}            
\setlength{\tabcolsep}{1.0mm}
\renewcommand{\footnoterule}{}                          
\begin{tabular}{c c c c c c c c c c c c}        
\hline \hline
\noalign{\smallskip}
$ \rm BJD_{UTC}$ & FWHM & $\pm 1~\sigma$ & BIS & $\pm 1~\sigma$ & $S_\mathrm{MW}$  & $\pm 1~\sigma$ & H$\alpha$ & $\pm 1~\sigma$ & $\log R^\prime_\mathrm{HK}$  & $\pm 1~\sigma$ & Case  \\
$-2400000$   & ($\ms$) & ($\ms$) & ($\ms$) & ($\ms$) &  &  &   &  &   &   &   \\
\noalign{\smallskip}
\hline
\noalign{\smallskip}
 58416.564987  &   9217.38    &  1.22  &  $-24.72$  &   1.22  &  0.15276 &   0.00007 &  0.19409  &    0.00003 &  $-4.8780$  &  0.0005 &  PRE \\
 58420.623837  &   9224.85    &  1.96  &  $-25.61$  &   1.96  &  0.14156 &   0.00015 &  0.19536  &    0.00005 &  $-4.9609$  &  0.0012 &  PRE \\
 58421.615745  &   9227.77    &  0.73  &  $-22.07$  &   0.73  &  0.15821 &   0.00003 &  0.19463  &    0.00002 &  $-4.8428$  &  0.0002 &  PRE \\
 58426.662776  &   9227.79    &  0.71  &  $-22.60$  &   0.71  &  0.15796 &   0.00003 &  0.19356  &    0.00002 &  $-4.8443$  &  0.0002 &  PRE \\
 58432.512969  &   9228.28    &  0.84  &  $-22.20$  &   0.84  &  0.15594 &   0.00004 &  0.19424  &    0.00002 &  $-4.8571$  &  0.0003 &  PRE \\
 58459.511927  &   9223.40    &  0.77  &  $-22.31$  &   0.77  &  0.15720 &   0.00003 &  0.19401  &    0.00002 &  $-4.8491$  &  0.0002 &  PRE \\
 58466.533613  &   9229.44    &  0.94  &  $-21.07$  &   0.94  &  0.15445 &   0.00005 &  0.19307  &    0.00002 &  $-4.8668$  &  0.0003 &  PRE \\
 58477.521619  &   9221.53    &  1.15  &  $-22.32$  &   1.15  &  0.15084 &   0.00006 &  0.19498  &    0.00003 &  $-4.8912$  &  0.0004 &  PRE \\
 58488.535606  &   9224.84    &  0.72  &  $-23.49$  &   0.72  &  0.15771 &   0.00003 &  0.19420  &    0.00002 &  $-4.8459$  &  0.0002 &  PRE \\
 58490.544982  &   9226.34    &  1.02  &  $-26.12$  &   1.02  &  0.15389 &   0.00005 &  0.19423  &    0.00002 &  $-4.8705$  &  0.0003 &  PRE \\
 58686.921063  &   9230.05    &  0.73  &  $-20.53$  &   0.73  &  0.15562 &   0.00003 &  0.19371  &    0.00001 &  $-4.8592$  &  0.0002 & POST \\
 58688.776700  &   9229.43    &  1.32  &  $-20.95$  &   1.32  &  0.14594 &   0.00009 &  0.19404  &    0.00003 &  $-4.9266$  &  0.0006 & POST \\
 58688.788018  &   9226.33    &  1.33  &  $-20.75$  &   1.33  &  0.14713 &   0.00009 &  0.19414  &    0.00003 &  $-4.9177$  &  0.0006 & POST \\
 58706.867055  &   9230.44    &  0.64  &  $-20.97$  &   0.64  &  0.15624 &   0.00003 &  0.19359  &    0.00001 &  $-4.8552$  &  0.0002 & POST \\
 58719.844569  &   9233.17    &  0.56  &  $-20.03$  &   0.56  &  0.15645 &   0.00002 &  0.19260  &    0.00001 &  $-4.8539$  &  0.0001 & POST \\
 58721.863285  &   9230.29    &  0.59  &  $-21.67$  &   0.59  &  0.15586 &   0.00003 &  0.19183  &    0.00001 &  $-4.8576$  &  0.0002 & POST \\
 58725.809120  &   9233.34    &  1.07  &  $-25.90$  &   1.07  &  0.15119 &   0.00006 &  0.19240  &    0.00002 &  $-4.8887$  &  0.0004 & POST \\
 58731.758304  &   9227.36    &  0.89  &  $-22.07$  &   0.89  &  0.15447 &   0.00005 &  0.19196  &    0.00002 &  $-4.8667$  &  0.0003 & POST \\
 58737.814370  &   9229.18    &  0.76  &  $-21.81$  &   0.76  &  0.15532 &   0.00004 &  0.19181  &    0.00002 &  $-4.8611$  &  0.0002 & POST \\
 58741.624983  &   9230.72    &  0.67  &  $-22.56$  &   0.67  &  0.15650 &   0.00003 &  0.19167  &    0.00001 &  $-4.8535$  &  0.0002 & POST \\
 58753.617398  &   9229.38    &  0.57  &  $-20.43$  &   0.57  &  0.15774 &   0.00002 &  0.19201  &    0.00001 &  $-4.8457$  &  0.0001 & POST \\
 58759.665785  &   9226.86    &  0.87  &  $-21.22$  &   0.87  &  0.15529 &   0.00005 &  0.19140  &    0.00002 &  $-4.8613$  &  0.0003 & POST \\
 58761.674737  &   9231.98    &  0.89  &  $-22.14$  &   0.89  &  0.15507 &   0.00005 &  0.19134  &    0.00002 &  $-4.8627$  &  0.0003 & POST \\
 58766.630575  &   9227.68    &  0.73  &  $-21.08$  &   0.73  &  0.15627 &   0.00003 &  0.19349  &    0.00001 &  $-4.8550$  &  0.0002 & POST \\
 58768.626690  &   9230.88    &  0.67  &  $-20.20$  &   0.67  &  0.15632 &   0.00003 &  0.19225  &    0.00001 &  $-4.8547$  &  0.0002 & POST \\
 58771.609041  &   9225.63    &  0.92  &  $-20.52$  &   0.92  &  0.15308 &   0.00005 &  0.19140  &    0.00002 &  $-4.8759$  &  0.0003 & POST \\
 58773.688306  &   9229.26    &  0.76  &  $-19.84$  &   0.76  &  0.15537 &   0.00004 &  0.19123  &    0.00002 &  $-4.8608$  &  0.0002 & POST \\
 58776.572183  &   9229.53    &  0.77  &  $-23.01$  &   0.77  &  0.15689 &   0.00004 &  0.19341  &    0.00002 &  $-4.8510$  &  0.0002 & POST \\
 58776.694342  &   9226.76    &  1.02  &  $-21.14$  &   1.02  &  0.15571 &   0.00006 &  0.19374  &    0.00002 &  $-4.8586$  &  0.0004 & POST \\
 58778.631061  &   9228.84    &  0.63  &  $-21.75$  &   0.63  &  0.15701 &   0.00003 &  0.19376  &    0.00001 &  $-4.8503$  &  0.0002 & POST \\
 58780.645934  &   9232.92    &  0.53  &  $-20.98$  &   0.53  &  0.15806 &   0.00002 &  0.19356  &    0.00001 &  $-4.8437$  &  0.0001 & POST \\
 58782.560951  &   9230.00    &  0.78  &  $-22.72$  &   0.78  &  0.15610 &   0.00004 &  0.19400  &    0.00002 &  $-4.8561$  &  0.0003 & POST \\
 58784.531509  &   9231.58    &  1.13  &  $-23.46$  &   1.13  &  0.15248 &   0.00007 &  0.19473  &    0.00003 &  $-4.8799$  &  0.0005 & POST \\
 58786.600937  &   9230.66    &  0.60  &  $-22.13$  &   0.60  &  0.15859 &   0.00003 &  0.19447  &    0.00001 &  $-4.8404$  &  0.0002 & POST \\
 58788.597405  &   9230.31    &  0.67  &  $-21.94$  &   0.67  &  0.15782 &   0.00003 &  0.19385  &    0.00001 &  $-4.8452$  &  0.0002 & POST \\
 58792.609247  &   9232.60    &  0.75  &  $-23.67$  &   0.75  &  0.15683 &   0.00004 &  0.19428  &    0.00002 &  $-4.8514$  &  0.0002 & POST \\
 58801.619078  &   9231.18    &  0.72  &  $-21.68$  &   0.72  &  0.15837 &   0.00003 &  0.19497  &    0.00001 &  $-4.8418$  &  0.0002 & POST \\
 58811.617055  &   9226.77    &  0.98  &  $-23.98$  &   0.98  &  0.15492 &   0.00006 &  0.19406  &    0.00002 &  $-4.8638$  &  0.0004 & POST \\
 58815.611406  &   9228.68    &  0.67  &  $-20.22$  &   0.67  &  0.15829 &   0.00003 &  0.19547  &    0.00001 &  $-4.8423$  &  0.0002 & POST \\
 58817.547039  &   9228.72    &  0.66  &  $-22.56$  &   0.66  &  0.15879 &   0.00003 &  0.19577  &    0.00001 &  $-4.8392$  &  0.0002 & POST \\
 58819.554583  &   9235.31    &  1.30  &  $-21.66$  &   1.30  &  0.15002 &   0.00009 &  0.19524  &    0.00003 &  $-4.8969$  &  0.0006 & POST \\
\noalign{\smallskip}
\hline \hline
\label{tab:activ}
\end{tabular}
\end{table*}


\end{appendix}

\end{document}